\documentclass{article}
\usepackage[english]{babel}
\usepackage[T2A]{fontenc}

\usepackage[tbtags]{amsmath}
\usepackage{amsfonts,amssymb,mathrsfs,amscd, bm}
\usepackage{hyperref}
\usepackage{lscape}
\usepackage{tikz}
\usepackage{circuitikz}
\usepackage{caption}
\usepackage{subcaption}

\newcommand{\down}{\underline{\downarrow}}

\graphicspath{ {./images/} }

\numberwithin{equation}{section}
\begin{document}
    \title{Acoustic pulse propagation in a non-ideal shallow-water model}
    \author{A. Kaplun\footnote{University of Haifa, Israel,\,\,\href{mailto: a.kaplun.work@gmail.com}{a.kaplun.work@gmail.com}, \href{https://orcid.org/0000-0003-4666-7911}{ORCID: 0000-0003-4666-7911}}, B. Katsnelson\footnote{University of Haifa, Israel,\,\,\href{mailto: bkatsnels@univ.haifa.ac.il}{bkatsnels@univ.haifa.ac.il}}}
\date{}

    \maketitle

\begin{abstract}
This study develops a theoretical framework for modeling acoustic pulse propagation in a non-ideal shallow-water waveguide. We derive an $\varepsilon$-pseudo\-diffe\-ren\-tial operator ($\varepsilon$-PDO) formulation from the general three-dimensional wave equation, that accounts for vertical stratification, bottom interaction, and slow horizontal inhomogeneity. Using the operator separation of variables method and the WKB-ansatz, we obtain single-mode equations describing the evolution of amplitude and phase along rays. The approach incorporates non-self-adjoint operators to model energy leakage through the bottom and introduces a Hamiltonian formalism for eikonal and transport equations, enabling the computation of amplitude, time, and phase fronts. Analytical and numerical examples are provided for different boundary conditions, including Neumann (ideal), self-adjoint, and partially reflecting interfaces. The results extend previous semiclassical and ray-based theories of wave propagation by including dissipative effects and improving the physical realism of shallow-water acoustic modeling.\\

\textbf{Keywords:} shallow-water acoustics, ray tracing, semiclassical analysis, non-self-adjoint operators, dispersion\\
\textbf{MSC:} 34A30, 34A12,	34B24, 35L05, 35L20, 35S05, 35Q60, 70H05, 70H09, 76Q05, 81Q20
\end{abstract}
\subsubsection*{Acknowledgments}
We dedicate this work to the memory of Prof. Vasilii Mikhailovich Babich (1930 -- 2025), head of the St.Petersburg school in the theory of diffraction and wave propagation (PDMI RAS/LOMI and SPbU/LGU), whose foundational contributions shaped the modern asymptotic methods in wave theory \cite{belishev2021vasilii, 11263469}.

Aleksandr Kaplun is supported by the Center for Integration in Science, Israel Ministry of Aliyah and Integration and the post-doctoral scholarship at the Leon H. Charney School of Marine Sciences, University of Haifa.  
Boris Katsnelson is supported by the Israel Science Foundation, grant ISF-946/20. 
\section{Introduction}

Acoustic wave propagation in shallow-water environments is a fundamental problem in underwater acoustics, relevant to sonar system design, environmental monitoring, and marine geophysics. In such environments, the combination of variable bathymetry, stratified sound-speed profiles, and bottom losses leads to complex propagation effects that cannot be adequately described by idealized or purely self-adjoint models \cite{katsnelson2012fundamentals, etter2018underwater,jensen2011computational}. 

Classical formulations of underwater sound propagation rely on ray and mode theories \cite{weinberg1974horizontal, burridge2005horizontal, vcerveny2001seismic}, in which the acoustic field is represented as a superposition of horizontally propagating rays and vertically standing modes. The theoretical foundation for such methods was established in the 1970s-1980s through the development of the space-time ray method by Babich, Molotkov, and Buldyrev \cite{babich1998space, babich1980space}, and the horizontal ray formalism of Weinberg and Burridge \cite{weinberg1974horizontal}. These approaches were later generalized to complex and moving boundary problems \cite{vcerveny1982space, kirpichnikova1983reflection, babich1984complex, farra1989ray, farra1993ray,connor1974complex} and further extended into hybrid ray-mode formulations for inhomogeneous ducts \cite{felsen1981hybrid}. Together, these works established the Hamiltonian and Lagrangian foundations of modern acoustic and seismic wave propagation.

Subsequent advances in semiclassical and asymptotic analysis, notably the Maslov canonical operator approach \cite{maslov2001semi, kravtsov1990geometrical, kravtsov2012caustics, dobrokhotov2017new}, have provided powerful mathematical tools for describing wavefields in weakly inhomogeneous and dispersive media. These techniques, based on the theory of canonical transformations in phase space, yield compact asymptotic representations of solutions near caustics and turning points, where traditional WKB or geometric acoustics approximations fail. Further refinements of this framework have been developed in works on operator separation of variables \cite{belov2006operator} and modern integral representations of the Maslov operator \cite{dobrokhotov2017new}, leading to robust methods for analyzing adiabatic and non-adiabatic mode evolution in three-dimensional shallow-water waveguides \cite{petrov2019application}. 

Parallel developments in optical and electromagnetic pulse theory have introduced analogous Hamiltonian structures for dispersive and lossy media. The analysis of group velocity dispersion, pulse-front tilt, and angular dispersion in ultrashort optical pulses \cite{topp1975group, bor1985group, hebling1996derivation, hebling2008generation, osvay2004angular,kiselev2012nonparaxial,katsnelson2018variability,katsnel2012space} has strong mathematical correspondence with acoustic phase and amplitude evolution in horizontally varying environments. These analogies support the application of space-time ray methods and Hamiltonian formalisms beyond conservative systems, enabling cross-domain interpretations of wavefront dynamics in both acoustics and optics.

Building on this theoretical foundation, we construct a non-ideal model for acoustic pulse propagation in a shallow-water waveguide that incorporates geometric and frequency dispersion, as well as partial energy transmission through the seabed. Starting from the general three-dimensional wave equation, we derive an $\varepsilon$-pseudodifferential operator ($\varepsilon$-PDO) formulation that separates fast (vertical) and slow (horizontal and temporal) variables. Applying the operator separation of variables method \cite{belov2006operator} and the WKB-Maslov ansatz \cite{maslov2001semi}, we obtain eikonal and transport equations governing single-mode propagation within a Hamiltonian framework. This provides a physically consistent representation of mode dynamics and geometric spreading in non-ideal, weakly inhomogeneous environments.

The use of non-self-adjoint and pseudo-Hermitian operators has become increasingly important in the analysis of dissipative and open wave systems. In the context of underwater acoustics, such operators naturally arise when accounting for partial reflection and energy leakage at the seabed \cite{belov2006operator, petrov2019application}. Mathematically, this leads to the appearance of complex eigenvalues and biorthogonal mode structures \cite{bender1998real, mostafazadeh2002pseudo, moiseyev2011nonhermitian}, which can describe the attenuation and phase shift of propagating acoustic modes. The spectral properties of these non-self-adjoint systems differ fundamentally from those of Hermitian ones, as eigenfunctions are no longer orthogonal in the classical sense but form dual pairs associated with adjoint operators \cite{berry2004physics}. 

These theoretical developments have deep connections to semiclassical and Hamiltonian mechanics. The introduction of complex-valued Hamiltonians \cite{berry2004physics} allows one to model the evolution of wavefronts in media with dissipation and gain while preserving the geometric structure of phase space. Similar formalisms have been successfully applied in optical and quantum contexts to describe PT-symmetric and lossy systems, providing valuable mathematical tools for the analysis of non-ideal acoustic waveguides. Integrating these ideas into the $\varepsilon$-PDO and WKB frameworks presented here enables a unified treatment of energy transport, geometric dispersion, and dissipation within a single Hamiltonian formalism.

The resulting system provides explicit analytical and numerical expressions for acoustic fronts -- including amplitude, phase, and time fields -- under different bottom reflection conditions. This framework unifies and extends classical ray and mode theories, offering a consistent mathematical approach for describing non-ideal shallow-water acoustics and establishing a theoretical bridge between semiclassical analysis, dissipative wave mechanics, and ocean acoustics modeling.

\section{Wave equation}
\subsection{Initial problem}
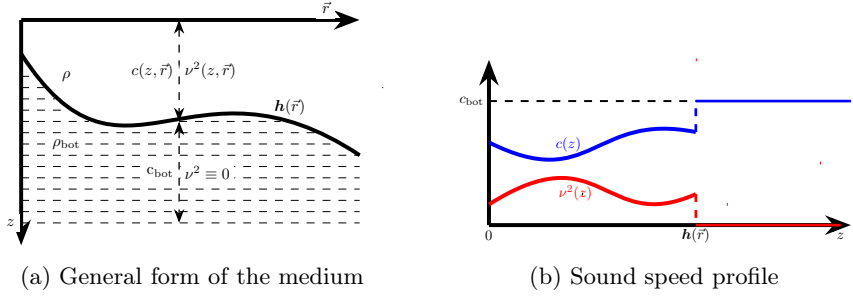
\begin{figure}

	\begin{subfigure}[b]{0.5\textwidth}
		\centering
		\begin{circuitikz}[scale = 0.6, transform shape]
			\tikzstyle{every node}=[font=\normalsize]
			\draw [line width=1.5pt, ->, >=Stealth] (2.5,11.25) -- (2.5,6.25)node[pos=0.9,left, fill=white]{$z  $};
			\draw [line width=1.5pt, ->, >=Stealth] (2.5,11.25) -- (10,11.25)node[pos=0.9,above, fill=white]{$\vec{r}$};
			\draw [line width=1.5pt, short] (2.5,10.5) .. controls (4.75,7) and (6.25,10.75) .. (10,8.25)node[pos = 0.85, above]{$\bm{h}(\vec{r})$};
			
			\draw [line width=0.2pt, dashed] (2.5,6.75) -- (10,6.75);
			\draw [line width=0.2pt, dashed] (2.5,7) -- (10,7);
			\draw [line width=0.2pt, dashed] (2.5,7.25) -- (10,7.25);
			\draw [line width=0.2pt, dashed] (2.5,7.5) -- (10,7.5);
			\draw [line width=0.2pt, dashed] (2.5,7.75) -- (10,7.75);
			\draw [line width=0.2pt, dashed] (2.5,8) -- (10,8);
			\draw [line width=0.2pt, dashed] (2.5,8.25) -- (10,8.25);
			\draw [line width=0.2pt, dashed] (2.5,8.5) -- (9.5,8.5);
			\draw [line width=0.2pt, dashed] (2.5,8.75) -- (9,8.75);
			\draw [line width=0.2pt, dashed] (2.5,9) -- (4,9);
			\draw [line width=0.2pt, dashed] (6,9) -- (8.25,9);
			\draw [line width=0.2pt, dashed] (2.5,9.25) -- (3.5,9.25);
			\draw [line width=0.2pt, dashed] (2.5,9.5) -- (3.25,9.5);
			\draw [line width=0.2pt, dashed] (2.5,9.75) -- (3,9.75);
			\draw [line width=0.2pt, dashed] (2.5,10) -- (2.75,10);
			\draw [line width=0.2pt, dashed] (10.5,9.75) -- (10.5,9.75);

			\draw [line width=0.5pt, <->, >=Stealth, dashed] (6,11.25) -- (6,9)node[pos=0.5,left, fill=white,rounded corners]{$c(z,\vec{r})$}node[pos=0.5,right, fill=white,rounded corners]{$\nu^2(z,\vec{r})$};
			\draw [line width=0.5pt, <->, >=Stealth, dashed] (6,9) -- (6,6.75)node[pos=0.5,left, fill=white,rounded corners]{$\mathrm{c}_{\mathrm{bot}}    $} node[pos=0.5,right, fill=white,rounded corners]{$\nu^2\equiv 0$};
			\node [font=\normalsize,fill=white,rounded corners] at (3.5,10) {$\rho$};
			\node [font=\normalsize,fill=white,rounded corners] at (3.5,8.5) {$\rho_{\mathrm{bot}}$};

		\end{circuitikz}
		
		\caption{General form of the medium}
		\label{medium}
	\end{subfigure}
	\begin{subfigure}[b]{0.5\textwidth}
		\centering
		\begin{circuitikz}[scale = 0.55, transform shape]
			\tikzstyle{every node}=[font=\normalsize]
			
			\draw [line width=0.75pt, >=Stealth, dashed] (1.25,11.75) -- (6.25,11.75)node[pos=0,left, fill=white]{$c_{\mathrm{bot}}$};
			\draw [line width=1.5pt, ->, >=Stealth] (1.25,8.75) -- (10,8.75);
			\node [font=\normalsize,rounded corners] at (9.75,8.5) {$z$};
			\draw [line width=1.5pt, ->, >=Stealth] (1.25,8.75) -- (1.25,12.75);
			\node [font=\normalsize,rounded corners] at (6.25,8.5) {$\bm{h}(\vec{r})$};
			\node [font=\normalsize,rounded corners] at (1.25,8.5) {$0$};
			\draw [ color={rgb,255:red,255; green,0; blue,0}, line width=1.5pt, short] (1.25,9.25) .. controls (4,11) and (4,8.5) .. (6.25,9.5)node[pos=0.35,below]{$\nu^2(z)$};
			\draw [line width=1pt, short] (3.5,9.5) -- (3.5,9.5);
			\draw [line width=1pt, short] (6.25,8.75) -- (9.75,8.75);
			\draw [ color={rgb,255:red,255; green,0; blue,0}, line width=1pt, short] (6.25,8.75) -- (9.75,8.75);
			\draw [ color={rgb,255:red,255; green,0; blue,0}, line width=1pt, short] (9.25,10.25) -- (9.25,10.25);
			\draw [ color={rgb,255:red,255; green,0; blue,0}, line width=1pt, dashed] (6.25,9.5) -- (6.25,8.75);
			\draw [ color={rgb,255:red,255; green,0; blue,0}, line width=1pt, dashed] (7,9.25) -- (7,9.25);

			\draw [ color={rgb,255:red,0; green,0; blue,255}, line width=1.5pt, short] (1.25,10.75) .. controls (3.75,9.5) and (3.75,11.5) .. (6.25,11)node[pos=0.35,above]{$c(z)$};
			\draw [ color={rgb,255:red,255; green,0; blue,0}, line width=1pt, dashed] (6.25,12.75) -- (6.25,12.75);
			\draw [ color={rgb,255:red,0; green,0; blue,255}, line width=1pt, dashed] (6.25,11) -- (6.25,11.75);
			\draw [ color={rgb,255:red,0; green,0; blue,255}, line width=1pt, short] (6.25,11.75) -- (10,11.75);

		\end{circuitikz}
		\caption{Sound speed profile}
		\label{sound speed}
	\end{subfigure}
	
	\caption{Medium properties}
	\label{medium full}
\end{figure}

Consider the wave equation describing the propagation of acoustic waves in the sea:
\begin{align}
	& \left[\frac{\partial ^2}{\partial x^2} + \frac{\partial ^2}{\partial y^2} + \frac{\partial ^2}{\partial z^2} - \frac{1}{c(z,x,y)^2} \frac{\partial ^2}{\partial t^2}\right] u(z,t,x,y) = 0.
\end{align}
Here, $x,y$ are horizontal coordinates and $z\ge 0$ is vertical, with $z=0$ being the equation of the water surface, $z= \bm{h}(x,y)$ being the equation of the lower surface (bottom) at the point $(x,y)$, and $c(z,x,y)$ being the speed of sound in the medium. The problem is defined in the half-space:
\begin{align}
	(x,y)\in \mathbb{R}^2,\quad z \in \overline{\mathbb{R}}_{+} = [0,+\infty). 
\end{align}
The desired function $u$ has the physical meaning of acoustic pressure. The sound speed  $c(x,y,z)$ satisfies the following properties:
\begin{align}
	& \mathbf{c}(x,y):= \left(||c^{-1}(\cdot,x,y)||_{L_{\infty}(\mathbb{R}_{+})}\right)^{-1},\quad   c(z,x,y) \ge\mathbf{c}(x,y) ,\\
	& c(z,x,y) \equiv {c_{\mathrm{bot}}}\equiv \mathrm{const},\quad z>\bm{h}(x,y),\,\, (x,y) \in \mathbb{R}^2.
\end{align}
Such notations allow us introduce more convenient function to use:
\begin{align}
	& \nu^2(z,x,y) :=\frac{c_{\mathrm{bot}}^2-c^2(z,x,y)}{c^2(z,x,y)} =  \frac{c_{\mathrm{bot}}^2}{c^2(x,y,z)}-1,\\ &\bm{\nu}^2(x,y) := ||\nu^2(\cdot,x,y)||_{L_{\infty}(\mathbb{R}_{+})} = \frac{\mathrm{c}_{\mathrm{bot}}^2 -\mathbf{c}^2(x,y) }{\mathbf{c}^2(x,y)} = \frac{\mathrm{c}_{\mathrm{bot}}^2 }{\mathbf{c}^2(x,y)} - 1,\\
	& 0 \le \nu(z,x,y) \le \bm{\nu}(x,y),\quad \nu(z,x,y) \equiv 0,\,\, z>\bm{h}(x,y), \,\,(x,y) \in \mathbb{R}^2.
\end{align}
Next, we describe the scale assumptions distinguishing the vertical coordinate 
$z$ from the slow variables. We say that all functions under our consideration depend on coordinates $t,x,y$ only via their combinations with some small parameter $\varepsilon\ll 1$: $\varepsilon t,\varepsilon x,\varepsilon y$. For example
\begin{align}
	& \nu(z,x,y) = \nu(z,\varepsilon x,\varepsilon y ),\quad \bm{h}(x,y)=\bm{h}(\varepsilon x,\varepsilon y).
\end{align}
Let us introduce the final change of variables, reflecting such properties along with transformation of $t$ to another "space variable":
\begin{align}
	& t \to \tau := \varepsilon\mathrm{c}_{\mathrm{bot}}t,\quad x,y \to \tilde{x},\tilde{y}:= \varepsilon x, \varepsilon y,\quad z \to z.
\end{align}
For simplicity, we omit the tilde and assume that $x,y$ are already scaled variables while the following notations are used:
\begin{align}
	& \bm{r}:=\left(\begin{array}{c}
		\tau\\
		\vec{r}
	\end{array}\right) = \left(\begin{array}{c}
		\tau\\
		x \\
		y
	\end{array}\right),\quad\bm{\nabla}:=\left(\begin{array}{c}
		\partial_{\tau}\\
		\vec{\nabla}
	\end{array}\right) = \left(\begin{array}{c}
		\partial_{\tau}\\
		\partial_{x} \\
		\partial_{y}
	\end{array}\right).
\end{align}
We now specify the boundary conditions. As the one on the top surface we would take Dirichlet one:
\begin{align}
	& u(z,\bm{r})|_{z=0} = 0.
\end{align}
Since we focus on waves propagating in the water, we impose the natural condition on the $z-$infinity for all possible other coordinates:
\begin{align}
	||u(\cdot,\bm{r})||^2_{L_2(\mathbb{R}_{+})} = \int\limits_{0}^{\infty} |u(z,\bm{r})|^2dz <\infty,
\end{align}
or, in shorter form, $u(\cdot,\bm{r}) \in L_2(\mathbb{R}_{+})$ for all $\bm{r}$. At the lower surface we would have transmission conditions of the following form:
\begin{align}
	u(z,\bm{r})|_{z = \bm{h}(\vec{r}) - }&= u(z,\bm{r})|_{z = \bm{h}(\vec{r}) + },\\
	\frac{d u}{dz}(z,\bm{r})|_{z = \bm{h}(\vec{r}) - } &= \alpha \frac{d u}{dz}(z,\bm{r})|_{z = \bm{h}(\vec{r}) + },\quad 0\le\alpha\le 1.
\end{align}
Here parameter $\alpha:= \frac{\rho}{\rho_{\mathrm{bot}}}$ represents relation between densities of the medium above and below the bottom (see Fig. \ref{medium full} )

\subsection{$\varepsilon-$PDO formulation}
The equation can now be written as:
\begin{align}\label{wave eq final form}
	& \hat{\mathcal{H}}(- i \varepsilon\bm{\nabla},- i \partial_z, \bm{r},z) u (\bm{r},z) = 0,
\end{align}
where
\begin{align}
	\notag &\hat{\mathcal{H}}	(- i \varepsilon\bm{\nabla},- i \partial_z, \bm{r},z):=\left\{\nu^2(z,\vec{r})+1\right\} (- i \varepsilon\partial_\tau)^2 - (-i\varepsilon\vec{\nabla})^2 -(- i \partial_z)^2. 
\end{align}
Let us define $\varepsilon$\textit{-pseudodifferential operator} ($\varepsilon$-PDO). We define direct and inverse (one-dimensional) \textit{ $\varepsilon-$Fourier transforms} $F_{s\to p_s}^{\varepsilon}$ and $F_{p_s\to s}^{\varepsilon}$  by the following equalities:
\begin{align}
	[F_{s\to p_s}^{\varepsilon}\chi(s)](p_s)&:= \frac{1}{\sqrt{2\pi i \varepsilon}}\int\limits_{\mathbb{R}_{s}}e^{- i\varepsilon^{-1} s p_s }\chi(s)ds,\\ 
	[F_{p_s\to s}^{\varepsilon}\tilde{\chi}(p_s)](s)&:= \frac{1}{\sqrt{-2\pi i \varepsilon}}\int\limits_{\mathbb{R}_{p_s}}e^{i\varepsilon^{-1} s p_s }\tilde{\chi}(p_s)dp_s.
\end{align}
Similarly we define  $F_{\bm{r}\to \bm{p}}^{\varepsilon}$ and $F_{\bm{p}\to \bm{r}}^{\varepsilon}$ as compositions of 3 corresponding transforms. Then, action of the $\varepsilon$-PDO $\hat{\mathcal{M}}(-i\varepsilon\bm{\nabla},\bm{r})$ with symbol $\mathcal{M}(\bm{p},\bm{r})$ is defined by the formula
\begin{align}
	\hat{\mathcal{M}}(-i\varepsilon\bm{\nabla},\bm{r})\chi(\bm{r})&:=F_{\bm{p}\to \bm{r}}^{\varepsilon}[\mathcal{M}(\bm{p},\bm{r})[F_{\bm{r}\to \bm{p}}^{\varepsilon}\chi(\bm{r})](\bm{p})](\bm{r}).
\end{align}
Hence, $\hat{\mathcal{H}}$ can be treated as an $\varepsilon$-PDO acting on \textit{slow} variables $\tau,x,y$ with $z$ as a \textit{fast} one. \textit{Operator-valued} symbol $\mathcal{H}$ of such operator is written in the form:
\begin{align}
	\notag&\mathcal{H}(\bm{p},- i \partial_z,\bm{r},z) :=\left\{-(- i \partial_z)^2+\nu^2(z,\vec{r}) p_{\tau}^2\right\}  +\left\{  p_{\tau}^2 -p_{x}^2 -  p_{y}^2 \right\}.
\end{align} 
Such form allows us to have the following decomposition:
\begin{align}
	\mathcal{H}(\bm{p},- i \partial_z,\bm{r},z) &=  \mathcal{L}(p_{\tau},\vec{r}) + \mathcal{H}_{\mathrm{class}}(\bm{p}),
\end{align}
where
\begin{align}
	\mathcal{L}(p_{\tau},\vec{r}) &:=\partial_z^2  +\nu^2(z,\vec{r}) p_{\tau}^2,\quad
	\mathcal{H}_{\mathrm{class}}(\bm{p}) := p_{\tau}^2 -p_{x}^2 -  p_{y}^2.
\end{align}
\subsection{WKB-ansatz for single mode}
We briefly outline the method for solving of the equation \ref{wave eq final form} using operator separation of the variables method. In details it can be found in work \cite{belov2006operator} but here we would focus on our exact problem. For simplification now we would describe single mode propagation ignoring mode coupling effects.

First step is to define expansion of Hamiltonian $\mathcal{H}$ with respect to small parameter $\varepsilon$. In our case we would have two separate cases - one (non-dissipative) with just zeroth order of expansion, in which after all separation of fast and slow variables, $\mathcal{H}$ does not depend on $\varepsilon$ explicitly, and the second - having term, proportional to $\varepsilon$, reflecting dissipation in the system. Also we can see, that our case is stationary in terms of \cite{belov2006operator}, because there is no dependence of Hamiltonian on some parameter along trajectories - time is considered as additional "space" variable.

The next step is separation of variables. The general form of the solution is:
\begin{align}
	& u(z,\bm{r},\varepsilon) = \hat{\bm{\psi}}(z,-i\varepsilon \bm{\nabla},\bm{r},\varepsilon) \left[\bm{A}(\bm{r},\varepsilon)\exp\{ i\varepsilon^{-1}\varphi(\bm{r})\}\right],
\end{align}
where $ \hat{ \bm{\psi}}$ is $\varepsilon$-PDO
with symbol $\bm{\psi} (z,p_{\tau},\vec{r},\varepsilon)$. Both $\bm{\psi}$ (mode) and $\bm{A}$ (amplitude) are assumed to be asymptotic series with respect to small parameter $\varepsilon$:
\begin{align}
	&\bm{\psi}(z,\bm{p},\bm{r},\varepsilon) = \sum_{j=0}^{\infty}\psi^j(z,\bm{p},\bm{r})\varepsilon^j,&&
	\bm{A}(\bm{r},\varepsilon) = \sum_{j=0}^{\infty}A^j(\bm{r})\varepsilon^j,
\end{align}
while $\varphi$ is viewed as phase, reflecting fast oscillations of the solution. We would also refer to the term $\bm{A}(\bm{r},\varepsilon)\exp\{ i\varepsilon^{-1}\varphi(\bm{r})\}$ as WKB-ansatz \cite{maslov2001semi}.

\subsection{$\varepsilon$-PDO and WKB-ansatz}
Let us discuss rather standard question - how we can express action of some $\varepsilon$-PDO $\hat{\mathcal{M}}(-i\varepsilon\bm{\nabla},\bm{r})$ on the WKB-ansatz. In such case we have:
\begin{align}
	\notag&\hat{\mathcal{M}}(-i\varepsilon\bm{\nabla},\bm{r})[\bm{A}(\bm{r},\varepsilon)\exp\{i\varepsilon^{-1}\varphi(\bm{r})\}](\bm{r}) =\\&= (2\pi \varepsilon)^{-3}\int\limits_{\mathbb{R}^3_{\bm{r}}\times\mathbb{R}^3_{\bm{p}}}\bm{A}(\bm{r}')\mathcal{M}(\bm{p},\bm{r})\exp\{i\varepsilon^{-1}(\varphi(\bm{r}') + \bm{p}\cdot(\bm{r}-\bm{r}'))\}d\bm{r}'d\bm{p}.
\end{align} 
As we are planning to use stationary phase method, we should define full phase $\bm{\Phi}(\bm{p},\bm{r}')$ as follows\footnote{Variable $\bm{r}$ is viewed as a parameter.}:
\begin{align}
	\Phi(\bm{p},\bm{r}') := \varphi(\bm{r}') + \bm{p}\cdot(\bm{r}-\bm{r}')).
\end{align}
Stationary point of this function is defined by equations
\begin{align}
	\bm{\nabla}_{\bm{p},\bm{r}'}\Phi= \left(\begin{array}{c}
		\bm{\nabla}_{\bm{p}}\Phi(\bm{p},\bm{r}')\\
		\bm{\nabla}_{\bm{r}'}\Phi(\bm{p},\bm{r}')
	\end{array}\right) = \left(\begin{array}{c}
		\bm{r}-\bm{r}'\\
		\bm{\nabla}\varphi(\bm{r}') - \bm{p}
	\end{array}\right) = 0.
\end{align}
This allows us to have simple condition $(\bm{p}_0,\bm{r}_0')=(\bm{\nabla}\varphi(\bm{r}),\bm{r})$. Hessian of phase $\Phi$ with respect to $\bm{p},\bm{r}'$ at this point is 
\begin{align}
	\bm{\nabla}^2_{\bm{p},\bm{r}'}\Phi  := \left(\begin{array}{cc}
		\bm{\nabla}_{\bm{p}}^2\Phi & \bm{\nabla}_{\bm{p}}\bm{\nabla}_{\bm{r}'}\Phi\\
		\bm{\nabla}_{\bm{p}}\bm{\nabla}_{\bm{r}'}\Phi & \bm{\nabla}_{\bm{r}'}^2\Phi 
	\end{array}
	\right)= \left(\begin{array}{cc}
		0_{3\times 3} & -\mathrm{I}_{3\times 3}\\
		-\mathrm{I}_{3\times 3} & \bm{\nabla}_{\bm{r}'}^2\varphi|_{\bm{r}'=\bm{r}} 
	\end{array}
	\right).
\end{align}
Since $\det |\bm{\nabla}^2_{\bm{p},\bm{r}'}\Phi| = 1$ regardless of the phase and also signature of this matrix is always 0, we can obtain expansion of the action of the $\varepsilon$-PDO  $\hat{\mathcal{M}}$ into series with respect to small parameter:
\begin{align}
	\notag &\hat{\mathcal{M}}(-i\varepsilon\bm{\nabla},\bm{r})[\bm{A}(\bm{r},\varepsilon)\exp\{i\varepsilon^{-1}\varphi(\bm{r})\}](\bm{r}) =\\\label{action of PDO} &= \exp\{i\varepsilon^{-1}\varphi(\bm{r})\}\left(\sum\limits_{k=0}^{\infty}\varepsilon^k\hat{\mathcal{M}}_k\right)\left(\sum\limits_{k=0}^{\infty}\varepsilon^k A^k\right)
\end{align}
We would mainly be focusing on first two terms of expansion in case of some exact operators, so let us present this two operator terms from decomposition of $\hat{\mathcal{M}}$:
\begin{align}
	\hat{\mathcal{M}}_0[A(\bm{r})]&:=\mathcal{M}(\bm{\nabla}\varphi(\bm{r}),\bm{r})A(\bm{r}),\\
	\hat{\mathcal{M}}_1[A(\bm{r})]&:=-i\left.\left\{ \left(\bm{\nabla}_{\bm{p}}\mathcal{M},\bm{\nabla}_{\bm{r}}A\right)+\frac{A}{2}\mathrm{tr}(\bm{\nabla}_{\bm{r}}^2\varphi \bm{\nabla}_{\bm{p}}^2\mathcal{M})\right\}\right|_{\bm{p}=\bm{\nabla}\varphi(\bm{r})}.
\end{align}
Finally, we write simplified form of \ref{action of PDO}: 
\begin{align}
	\notag \exp\{-i\varepsilon^{-1}\varphi(\bm{r})\}&\hat{\mathcal{M}}(-i\varepsilon\bm{\nabla},\bm{r})[\bm{A}(\bm{r},\varepsilon)\exp\{i\varepsilon^{-1}\varphi(\bm{r})\}](\bm{r}) =  \\\label{PDO on WKB} &=\hat{\mathcal{M}}_0[A^0] + \varepsilon\left\{\hat{\mathcal{M}}_0[A^1]+\hat{\mathcal{M}}_1[A^0]\right\}+ O(\varepsilon^2).
\end{align}

\section{Non-dissipative case}
\subsection{Real eigenvalues}
To apply the method of \cite{belov2006operator, maslov2001semi}, we study the following eigenproblem in the space $L_2(\mathbb{R}_+)$ with boundary conditions, described above:
\begin{align}\label{eigenproblem}
	& \mathcal{H}(\bm{p},-i\partial_z,\bm{r},z)\psi^0(z,\bm{p},\bm{r}) = \mathcal{H}_{\mathrm{eff}}(\bm{p},\bm{r}) \psi^0(z,\bm{p},\bm{r}).
\end{align} 
Let us recall that we have the following decomposition of the Hamiltonian:
\begin{align}
	\mathcal{H}(\bm{p},- i \partial_z,\bm{r},z) &=  \mathcal{L}(p_{\tau},\vec{r}) + \mathcal{H}_{\mathrm{class}}(\bm{p}).
\end{align}
For eigenproblem \ref{eigenproblem} we have, following general horizontal-vertical decomposition ideas \cite{burridge2005horizontal,weinberg1974horizontal}:
\begin{align}
	&\mathcal{H}_{\mathrm{eff}}(\bm{p},\bm{r}):=\lambda(p_{\tau},\vec{r})+\mathcal{H}_{\mathrm{class}}(\bm{p}) =\lambda(p_{\tau},\vec{r})+  p_{\tau}^2 - p_x^2 - p_y^2,\\
	& \mathcal{L}(p_{\tau},\vec{r}) \psi^0(z,\bm{p},\bm{r}) =\lambda(p_{\tau},\vec{r}) \psi^0(z,\bm{p},\bm{r}),\quad ||\psi^0||_{L_2(\mathbb{R}_+)}=1.
\end{align}
We introduce an assumption to simplify subsequent equations. We assume, that, even though operators $\mathcal{L}(p_{\tau},\vec{r}) $ are, generally speaking, non-self-adjoint, we would need the following to be true~\cite{bender1998real,mostafazadeh2002pseudo}:
\begin{align}
	\label{assumption IM 0}& \Im \mathcal{H}_{\mathrm{eff}}(\bm{p},\bm{r}) = \Im \lambda(p_{\tau},\vec{r}) = 0.
\end{align}
The functions $\bm{A}$ and $\varphi$ then satisfy the equation:
\begin{align}\label{dispersion relation}
	&\hat{L}(-i\varepsilon\bm{\nabla},\bm{r}, \varepsilon)\left[\bm{A}(\bm{r},\varepsilon)\exp\{ i\varepsilon^{-1}\varphi(\bm{r})\}\right] = 0,
\end{align}
where $\hat{L}$ is $\varepsilon$-pseudodifferential operator with symbol
\begin{align}
	L(\bm{p},\bm{r}, \varepsilon) = \sum_{j=0}^{\infty}L^j(\bm{p},\bm{r})\varepsilon^j.
\end{align}
Equation \ref{dispersion relation} we would also call \textit{quantization of the dispersion relation} as it was in work \cite{belov2006operator}. It can be shown, that first term has form
\begin{align}
	L^0(\bm{p},\bm{r}) = \mathcal{H}_{\mathrm{eff}}(\bm{p},\bm{r}) = \lambda_l(p_{\tau},\vec{r})+\mathcal{H}_{\mathrm{class}}(\bm{p}).
\end{align}
Here we can clearly see the main reason for decomposition of the Hamiltonian. Now, each pseudodifferential operator  $L^0$ consists of two parts: purely differential, corresponding to $\mathcal{H}_{\mathrm{class}}(\bm{p})$ and invariant for each problem under our consideration, and pseudodifferential, generated by eigenvalues $\lambda(p_{\tau},\vec{r})$ \cite{dobrokhotov2017new}. 

We next derive the equations defining $\psi^1$ and $L^1$. From the general theory we have
\begin{align}
	& \left[\mathcal{H}(\bm{p},-i\partial_z,\bm{r},z) -  \mathcal{H}_{\mathrm{eff}}(\bm{p},\bm{r})\right]\psi^1 = \hat{\mathcal{D}}\psi^0 + \psi^0 L^1,
\end{align}
where
\begin{align}
	&\mathcal{H}(\bm{p},-i\partial_z,\bm{r},z) -  \mathcal{H}_{\mathrm{eff}}(\bm{p},\bm{r}) = \mathcal{L}(p_{\tau},\vec{r})-k^2(p_{\tau},\vec{r}),\\
	&\hat{\mathcal{D}} = 2 i \left[(\nu^2+1)p_{\tau}\partial_{\tau}\cdot- (\vec{p},\vec{\nabla} \cdot) - \left(\frac{1}{2}\frac{\partial \lambda}{\partial \vec{p}},\frac{\partial}{\partial \vec{p}}\cdot\right)\right],
\end{align}
while $L^1$ is defined from orthogonality condition:
\begin{align}
	L^1(\bm{p},\bm{r})&:=-\frac{\langle\hat{\mathcal{D}}\psi^0,\xi^0\rangle}{\langle\psi^0,\xi^0\rangle}  ={2i} \left(\vec{p},\frac{\langle\vec{\nabla}\psi^0,\xi^0\rangle}{\langle\psi^0,\xi^0\rangle}\right),
\end{align}
where $\xi^0$ is eigenfunction of the corresponding conjugated operator $(L^0)^*$. Then equation for the term $\psi^1$ has form
\begin{align}
	& \left[\mathcal{L}(p_{\tau},\vec{r})-\lambda(p_{\tau},\vec{r})\right] \psi^1 =2 i\left(\vec{p},\psi^0\frac{\langle\vec{\nabla} \psi^0,\xi^0\rangle}{\langle\psi^0,\xi^0\rangle}-\vec{\nabla}\psi^0\right). 
\end{align}
From this form, $\psi^1$  is independent of the time parameter $\tau$.

\subsection{Equation for WKB-ansatz}
We now derive equations for amplitude and phase. We start from  \ref{dispersion relation} and expand all objects with respect to small parameter up to the term $O(\varepsilon^2)$
\begin{align}
	\notag&\left\{\hat{\mathcal{H}}_{\mathrm{eff}}+ \varepsilon\hat{L}^1 +O(\varepsilon^2)\right\}[\bm{A}(\bm{r},\varepsilon)\exp\{i\varepsilon^{-1}\varphi(\bm{r})\}] = O(\varepsilon^2).
\end{align}
Applying previous results for expansion \ref{PDO on WKB}, we have (omitting universal exponential factor) 
\begin{align}
	[\hat{\mathcal{H}}_{\mathrm{eff}}]_0[A^0] &+ \varepsilon\left\{[\hat{\mathcal{H}}_{\mathrm{eff}}]_0[A^1] + [\hat{\mathcal{H}}_{\mathrm{eff}}]_1[A^0] + \hat{L}^1_0[A^0]\right\} = O(\varepsilon^2)
\end{align}
It appears to be convenient to change some notations, related to the Hamiltonian itself. Let us define the following objects:
\begin{align}
	\bm{H}(\bm{p},\bm{r})	&: = \frac{p_\tau^2+\lambda(p_{\tau},\vec{r}) - p_x^2-p_y^2}{2},\\
	\left(\begin{array}{c}
		\bm{k}(\bm{p},\bm{r})	\\
		\bm{l}(\bm{p},\bm{r})
	\end{array}\right)&: =  \left(\begin{array}{c}
		\bm{\nabla}_{\bm{p}}\bm{H}(\bm{p},\bm{r}) 	\\
		\bm{\nabla}_{\bm{r}}\bm{H}(\bm{p},\bm{r})
	\end{array}\right) = \left(\begin{array}{c}
		p_{\tau} + \frac{1}{2}\frac{\partial \lambda}{\partial p_{\tau}}	\\
		-\vec{p} \\
		0\\
		\frac{1}{2}\vec{\nabla}\lambda
	\end{array}\right).
\end{align}
Explicit forms corresponding decompositions of each operator are:
\begin{align}
	[\hat{\mathcal{H}}_{\mathrm{eff}}]_0[A(\bm{r})]&=\mathcal{H}_{\mathrm{eff}}(\bm{\nabla}\varphi(\bm{r}),\bm{r})A(\bm{r})=2\bm{H}(\bm{\nabla}\varphi(\bm{r}),\bm{r})A(\bm{r}),\\
	[\hat{\mathcal{H}}_{\mathrm{eff}}]_1[A(\bm{r})]&=-i A^{-1}(\bm{r})\bm{\nabla}_{\bm{r}}\cdot\left\{A^2(\bm{r})\bm{k}(\bm{\nabla}\varphi(\bm{r}),\bm{r})\right\},\\ 
	\hat{L}_0^1[A(\bm{r})] &= {2i} \left(\vec{\nabla}\varphi,\frac{\langle\vec{\nabla}\psi^0,\xi^0\rangle}{\langle\psi^0,\xi^0\rangle}\right) A(\bm{r})
\end{align}
From here, we obtain following two equations, in accordance with semiclassical WKB formalism \cite{maslov2001semi,kravtsov1990geometrical}:
\begin{align}
	\label{eikonal equation} \bm{H}(\bm{\nabla}\varphi(\bm{r}),\bm{r}) &= 0,\\
	\label{transport equation} \bm{\nabla}_{\bm{r}}\cdot\left\{[A^0(\bm{r})]^2\bm{k}(\bm{\nabla}\varphi(\bm{r}),\bm{r})\right\}&=2[ A^0 (\bm{r})]^2\left(\vec{\nabla}\varphi,\frac{\langle\vec{\nabla} \psi^0,\xi^0\rangle}{\langle\psi^0,\xi^0\rangle}\right) .
\end{align}
Naturally, the first one (\ref{eikonal equation}) we would call \textit{eikonal equation}, second one (\ref{transport equation}) - \textit{transport equation}.

\section{Vertical modes}
We now discuss the structure of the operators $\mathcal{L}(p_{\tau},\vec{r})$, their eigenvalues $\lambda$ and eigenfunctions $\psi$, as well as conjugated ones. 

\subsection{Eigenvalue-dependent problem}
As shown later, the key to constructing vertical modes is the solutions of the following problem on the interval $[0, 1]$:
\begin{align}
	&  \left[\frac{d^2}{d s^2} +\bm{w}^2(s,p_{\tau},\vec{r})  - k^2\right] \bm{\psi}(s,p_{\tau},\vec{r},k) = 0,\\
	& \bm{\psi}(s)|_{s = 0} = 0,\quad	\frac{d \bm{\psi}}{ds}(s)\bigg|_{s = 1} =- \alpha k,\quad \bm{\psi}(s)|_{s = 1} = 1,\\
	&\bm{w}^2(s,p_{\tau},\vec{r}):=\nu^2(s \bm{h}(\vec{r}),\vec{r})  p^2_{\tau}\bm{h}^2(\vec{r})
\end{align}
Such problem is stated for the pair: eigenvalue\footnote{More precise would be to call $k^2$ eigenvalues, but for simplification of the text we would call $k$ eigenvalues too.} $k$ and eigenfunction $\bm{\psi}(s,p_{\tau},\vec{r},k)$, corresponding to this eigenvalue (set of eigenvalues  $k$ we denote with $\sigma_d(p_{\tau},\vec{r})$). Since the problem depends strongly on these parameters, we impose the following assumptions on its spectral properties. Let us introduce notation:
\begin{align}
	\sigma_d^+(p_{\tau},\vec{r}):= \{k\,|\,k \in \sigma_d(p_{\tau},\vec{r}), \Re k>0  \} 
\end{align}
For this set, we assume that:
\begin{itemize}
	\item each eigenvalue is simple (has only one eigenfunction corresponding to it);
	\item each eigenvalue depends smoothly on parameters $p_{\tau},\vec{r}$, while they vary in some area $\Omega\subset\mathbb{R}^3$;
	\item eigenvalues are separated from one another: there exists some constant $\delta(\Omega)>0$, corresponding to the whole area $\Omega$, such that
	\begin{align}
		& \inf _{\begin{array}{c}
				k \in \sigma_d^+(p_{\tau},\vec{r}),\\
				(p_{\tau},\vec{r})\in \Omega
		\end{array} }\mathrm{dist}\left(k,\sigma_d^+ (p_{\tau},\vec{r})\right)\setminus \{k\})>\delta(\Omega),
	\end{align}
\end{itemize}
For complex eigenvalues and eigenfunctions we have obvious relation:
\begin{align}
	&k \in \sigma_d^+(p_{\tau},\vec{r}) \Rightarrow \overline{k} \in \sigma_d^+(p_{\tau},\vec{r}), && 
	\bm{\psi}(s,p_{\tau},\vec{r},\overline{k}) = \overline{\bm{\psi}}(s,p_{\tau},\vec{r},k).
\end{align}
\subsection{Operators $ \mathcal{L}(p_{\tau},\vec{r})$ }
Now, let us define family of operators, depending on $p_{\tau},\vec{r}$ as parameters:
\begin{align}
	& \mathcal{L}(p_{\tau},\vec{r}):L_2(\mathbb{R}_{+}) \to L_2(\mathbb{R}_{+}),\quad \mathcal{L}(p_{\tau},\vec{r})\psi:= \left[\frac{d^2}{d z^2} +\nu^2(z,\vec{r}) p_{\tau}^2\right] \psi
\end{align}
with domains $D(\mathcal{L}(p,\vec{r}))\subset L_2(\mathbb{R}_{+})$ (for each point $p,\vec{r}$) defined by conditions:
\begin{align}
	&\psi(z)|_{z=0}= 0,\quad
	\left.\left(\begin{array}{c}
		\psi\\
		\frac{d \psi}{dz}
	\end{array}\right)\right|_{z = \bm{h}(\vec{r}) - } = \left(\begin{array}{cc}
		1 & 0\\
		0 & \alpha
	\end{array}\right) \left.\left(\begin{array}{c}
		\psi\\
		\frac{d \psi}{dz}
	\end{array}\right)\right|_{z = \bm{h}(\vec{r}) + }
\end{align}
Eigenvalues of such operators are defined as follows:
\begin{align}
	\mathcal{L}(p_{\tau},\vec{r})\psi = \lambda \psi
\end{align}
From the conditions $\psi\in L_2(\mathbb{R}_{+})$ and $\nu^2(z,\vec{r})\equiv 0$ for $z> \bm{h}(\vec{r})$ we have form of any eigenfunction below $ \bm{h}(\vec{r})$ (this part of function we would denote $\psi^{\down}$):
\begin{align}
	\psi^{\down}(z,\lambda)  =  c\exp \left(-k  \left[\frac{z}{\bm{h}(\vec{r})}-1\right]\right),
\end{align}
where $k:=\bm{h}(\vec{r})\sqrt{\lambda}, \Re k>0$.
As a result, one can see, that such form of the solution bellow the bottom leads to the reduction of our problem in the whole $L_2(\mathbb{R}_+)$ to the one, we studied above. Let us introduce the following notation
\begin{align}
	\bm{\beta}(\alpha,p_{\tau},\vec{r},k):=&\left[||\bm{\psi}||^2(p_{\tau},\vec{r},k) + \frac{\alpha^2}{2\Re k (p_{\tau},\vec{r})}\right]^{-\frac{1}{2}},\\
	\notag\Psi(z,\alpha,p_{\tau},\vec{r},k):=&\frac{\bm{\beta}(\alpha, p_{\tau},\vec{r},k)}{\sqrt{\bm{h}(\vec{r})}}   \times \\ & \times\left\{\begin{array}{ll}
		\bm{\psi}\left(\frac{z}{\bm{h}(\vec{r})},p_{\tau},\vec{r},k\right)	,& z\in[0,\bm{h}(\vec{r})],\\[1ex]
		\alpha\exp \left(-k  \left[\frac{z}{\bm{h}(\vec{r})}-1\right]\right), & z>\bm{h}(\vec{r}),
	\end{array}\right..
\end{align}
We should mention, that for function $\bm{\psi}$ parameter $\alpha$ is \textit{always} included and dependence in $\bm{\beta}$ and $\Psi$ on  $\alpha$ is \textit{only} through the coefficients. Then function $\Psi(z,1,p_{\tau},\vec{r},k)$ is normalized eigenfunction of the operator $\mathcal{L}(p_{\tau},\vec{r})$ corresponding to the eigenvalue $\lambda = \frac{k^2}{\bm{h}^2(\vec{r})}$. Since operator is not self-adjoint, such functions are not necessary orthogonal (the same would also be for eigenfunctions of the adjoint one).
\subsection{Operators $ \mathcal{L}^*(p,\vec{r})$ }
We now describe the adjoint operators $\mathcal{L}^*(p,\vec{r})$:
\begin{align}
	& \mathcal{L}^*(p_{\tau},\vec{r}):L_2(\mathbb{R}_{+}) \to L_2(\mathbb{R}_{+}),\quad \mathcal{L}^*(p_{\tau},\vec{r})\xi:= \left[\frac{d^2}{d z^2} +\nu^2(z,\vec{r}) p_{\tau}^2\right] \xi
\end{align}
with domains $D(\mathcal{L}^*(p,\vec{r}))\subset L_2(\mathbb{R}_{+})$  defined by conditions:
\begin{align}
	&\xi(z)|_{z=0}= 0,\quad
	\left(\begin{array}{cc}
		\alpha  & 0\\
		0 & 1
	\end{array}\right)	\left.\left(\begin{array}{c}
		\xi\\
		\frac{d \xi}{dz}
	\end{array}\right)\right|_{z = \bm{h}(\vec{r}) - } =  \left.\left(\begin{array}{c}
		\xi\\
		\frac{d \xi}{dz}
	\end{array}\right)\right|_{z = \bm{h}(\vec{r}) + }
\end{align}
Thus, the operators $\mathcal{L}(p,\vec{r})$ are self-adjoint if and only if $\alpha=1$, which is not general case.

Using similar arguments and same notations as in the case of operators  $\mathcal{L}(p,\vec{r})$ we obtain that functions $\Psi(z,\alpha,p_{\tau},\vec{r},k^*)$ are normalized eigenfunctions of the operator $\mathcal{L}^*(p_{\tau},\vec{r})$ corresponding to eigenvalues $\lambda^* = \frac{(k^*)^2}{\bm{h}^2(\vec{r})}$.
\subsection{Relation between operators}
Let us discuss relations between operators, described above. From the general theory we know that for each eigenvalue $k$ of the operator $\mathcal{L}$ we have corresponding eigenvalue $k^* := \overline{k}$ of the operator $\mathcal{L}^*$. Introducing notation
\begin{align}
	\sigma_d (\mathcal{L})&:= \left\{\frac{k_l^2}{\bm{h}^2(\vec{r})}\right\}_{l=0}^{n(p_{\tau},\vec{r})}, \quad \Re k_0> \Re k_1> \dots > \Re k_{n(p_{\tau},\vec{r})}>0,\\
	\sigma_d (\mathcal{L}^*)&:= \left\{\frac{(k^*)^2}{\bm{h}^2(\vec{r})}\right\}_{l=0}^{n(p_{\tau},\vec{r})}, \quad \Re k_0^*> \Re k_1^*> \dots > \Re k_{n(p_{\tau},\vec{r})}^*>0,
\end{align} 
where $k^*_l = \overline{k}_l$. 

Let us additionally describe simplifications for the real $k_l=k_l^*\in\mathbb{R}_+$ and real-valued functions $\Psi_l$, corresponding to them:
\begin{align}
	\Psi_l(z,p_{\tau},\vec{r}):= \Psi(z,p_{\tau},\vec{r},k_l),\quad \bm{\psi}_l(s,p_{\tau},\vec{r}):= \bm{\psi}(s,p_{\tau},\vec{r},k_l).
\end{align}
In such conditions we have simple formula scalar product $\langle\Psi_l(1),\Psi_l(\alpha)\rangle$:
\begin{align}
	\notag\langle\Psi_l(1),\Psi_l(\alpha)\rangle =& \frac{\bm{\beta}(\alpha)}{\bm{\beta}(1)} + \frac{\alpha - 1}{\sqrt{2 k ||\bm{\psi}||^2 +1 }\sqrt{2 k ||\bm{\psi}||^2 +\alpha^2 }}
\end{align}
and also for the following fraction
\begin{align}
	\notag\frac{\langle\vec{\nabla}\Psi_l(1),\Psi_l(\alpha)\rangle}{\langle\Psi_l(1),\Psi_l(\alpha)\rangle} =& \frac{\alpha - 1}{\alpha + 2 k ||\bm{\psi}||^2} \times\left[  k \frac{\vec{\nabla} \bm{h}}{\bm{h}} - \frac{\vec{\nabla} \{k||\bm{\psi}||^2\}}{2 k ||\bm{\psi}||^2 +1}\right]. 
\end{align}

\section{Hamiltonian dynamics}
\subsection{Eikonal equation}
We examine \ref{eikonal equation} in detail: 
\begin{align}
	\label{eikonal equation 2}    &  \frac{1}{2} \left[\left(\frac{\partial \varphi(\bm{r})}{\partial \tau}\right)^2 +  \lambda\left(\frac{\partial \varphi(\bm{r})}{\partial \tau},\vec{r}\right)-\left(\vec{\nabla}\varphi(\bm{r})\right)^2\right]= 0.
\end{align}
We solve it using the standard method of characteristics applied to the Hamiltonian $\bm{H}$ \cite{babich1998space,kravtsov1990geometrical,maslov2001semi}.
Let us introduce phase space $\mathbb{R}^3_{\bm{r}}\times \mathbb{R}^3_{\bm{p}}$ with its elements having the following form:
\begin{align}
	& \bm{f} := \left(\begin{array}{c}
		\bm{r}\\
		\bm{p}
	\end{array}\right) =  \left(\begin{array}{c}
		\tau\\
		\vec{r}\\
		p_\tau\\
		\vec{p}
	\end{array}\right) \in  \mathbb{R}^3_{\bm{r}}\times \mathbb{R}^3_{\bm{p}},&& \bm{\nabla}_{\bm{f}}:= \left(\begin{array}{c}
		\bm{\nabla}_{\bm{r}}\\
		\bm{\nabla}_{\bm{p}}
	\end{array}\right).
\end{align}
Then characteristics of equation \ref{eikonal equation 2} are solutions  of system
\begin{align}
	& \frac{d \bm{f}}{d\sigma} =\bm{J} \bm{\nabla}_{\bm{f}} \bm{H}, && \bm{J} := \mathrm{J}\otimes \mathrm{I}_{3\times 3},\quad \mathrm{J}:= \left(\begin{array}{cc}
		0 & 1 \\
		-1 & 0
	\end{array}\right),
\end{align} 
or, in more explicit form, 
\begin{align}
	\label{Ham syst expl}& \frac{d }{d\sigma}\left(\begin{array}{c}
		\bm{r}\\
		\bm{p}
	\end{array}\right) =\left(\begin{array}{cc}
		0  & \mathrm{I}_{3\times 3} \\
		-\mathrm{I}_{3\times 3}  & 0
	\end{array}\right) \left(\begin{array}{c}
		\bm{\nabla}_{\bm{r}} \bm{H}  \\
		\bm{\nabla}_{\bm{p}} \bm{H}
	\end{array}\right) = \left(\begin{array}{c}
		\bm{k} \\
		-\bm{l}
	\end{array}\right).
\end{align} 
Here $\mathrm{J}$ is standard $2\times 2$ symplectic matrix and $\otimes$ is Kronecker  product of matrices. Parameter $\sigma$ does not have any particular physical meaning at this moment and represents some variable along characteristic. 

We next specify the initial conditions for this problem. Let us introduce 2-dimensional manifold $\Lambda_0$ in phase space:
\begin{align}
	&\Lambda_0:=\{(\bm{r}_0(\mu),\bm{p}_0(\mu))|\,\,\mu \in \Omega\} \subset \mathbb{R}^3_{\bm{r}}\times \mathbb{R}^3_{\bm{p}},\quad\mu \in \Omega \subset \mathbb{R}^2.
\end{align}
For simplicity, we assume that is $C^{\infty}$-smooth. On this manifold we introduce function $\varphi_0(\mu):=\varphi_0(\bm{r}_0(\mu))$. Then, to have correctly defined Cauchy problem, we need to have the following relations and conditions satisfied for all $\mu \in \Omega$:
\begin{align}
	& \bm{\nabla}_{\bm{r}}\varphi_0(\bm{r}_0(\mu)) = \bm{p}_0(\mu) ,&
	& \bm{k}(\bm{p}_0(\mu),\bm{r}_0(\mu))\neq 0, &
	& \mathrm{rank} \frac{\partial \bm{r}_0(\mu)}{\partial \mu} = 2.
\end{align} 
This manifold $\Lambda_0$, and all ray-shifted ones $\Lambda_\sigma$, are Lagrangian:
\begin{align}
	&\Lambda_\sigma:=\{(\bm{r}(\sigma,\mu),\bm{p}(\sigma,\mu))|\,\,\mu \in \Omega\},
\end{align}
where $\bm{r}(\sigma,\mu),\bm{p}(\sigma,\mu)$ are solutions of system \ref{Ham syst expl}  with initial values $\bm{r}_0(\mu),\bm{p}_0(\mu)$. Trajectories $\bm{f}(\sigma)$ in phase space we would call characteristics, while their projections $\bm{r}(\sigma)$ - rays. From the general theory we know that  trajectories $\bm{f}(\cdot,\mu)$ and $\bm{f}(\cdot,\mu')$ do not intersect, while corresponding rays can (and will, forming caustics). Pairs $\mathfrak{r}:=(\sigma,\mu)$ we would naturally call \textit{ray-centered coordinates} or simply \textit{ray coordinates}. For more detailed description we need to introduce notations for Jacobi matrices:
\begin{align}
	&\bm{f}_{\mathfrak{r}}(\mathfrak{r}) :=  \frac{\partial \bm{f}}{\partial \mathfrak{r}}(\mathfrak{r}) = \left(\begin{array}{c}
		\frac{\partial \bm{r}}{\partial \mathfrak{r}}(\mathfrak{r})\\
		\frac{\partial \bm{p}}{\partial \mathfrak{r}}(\mathfrak{r})
	\end{array}\right)=:\left(\begin{array}{c}
		\bm{r}_{\mathfrak{r}}(\mathfrak{r})\\
		\bm{p}_{\mathfrak{r}}(\mathfrak{r})
	\end{array}\right)\in \mathbb{M}^{6\times 3} 
\end{align}
Then general facts above can be expressed in such form:
\begin{align}
	& \det |\bm{r}_{\mathfrak{r}}(0,\mu)|\neq 0, && \mathrm{rank}\,\bm{f}_{\mathfrak{r}}(\sigma,\mu) =  \mathrm{rank}\,\bm{f}_{\mathfrak{r}}(0,\mu) = 3.
\end{align}
Solution $\varphi$ of \ref{eikonal equation 2} for given characteristic $\bm{f}(\sigma,\mu)$ can be be found via formula:
\begin{align}
	\notag \varphi(\sigma,\mu) &= \varphi_0(\mu) + \int\limits_{0}^{\sigma} (\bm{p}(\sigma',\mu),\bm{k}(\sigma',\mu))d\sigma' =\\& =\varphi_0(\mu) + \int\limits_{0}^{\sigma}p_{\tau}(\sigma',\mu)\left[\frac{1}{2}\frac{\partial {\lambda}}{\partial p_{\tau}}(\sigma',\mu) - \frac{{\lambda}(\sigma',\mu)}{p_{\tau}(\sigma',\mu)}\right]d\sigma'  .
\end{align}

\subsection{Variation equations}
To study details of wave propagation beyond ray paths, we use variational methods for Hamilton systems \cite{farra1993ray,farra1989ray}. We define variation of some function $g(\sigma,\mu)$ of ray coordinates as follows:
\begin{align}
	& \Delta[g](\mathfrak{r},\delta\mu):= g(\sigma,\mu+\delta\mu) - g(\sigma,\mu).
\end{align}
This allows us to write the following equations for such variations in case of $g(\mathfrak{r}) = \bm{f}(\mathfrak{r})$:
\begin{align}
	\label{first var syst}& \left\{\begin{array}{l}
		\frac{d }{d\sigma}\Delta[\bm{f}] =\bm{J} \bm{\nabla}_{\bm{f}}^2 \bm{H}\Delta[\bm{f}] \\[1ex]
		\Delta[\bm{f}]|_{\sigma =0 }= \Delta_0[\bm{f}]
	\end{array}\right. .
\end{align}
It appears more convenient to study this system via the propagator matrix (also known as the monodromy matrix) $\bm{P}(\mathfrak{r})$ defined by the following system:
\begin{align}
	\label{propagator syst}& \left\{\begin{array}{l}
		\frac{d }{d\sigma}\bm{P} =\bm{J}\bm{\nabla}_{\bm{f}}^2 \bm{H}\bm{P}\\[1ex]
		\bm{P}|_{\sigma = 0} = \mathrm{I}_{6\times 6}
	\end{array}\right..
\end{align}
Any propagator matrix defined by the system above is symplectic, i.e. :
\begin{align}
	& \bm{P}^t\bm{J}\bm{P} =\bm{J}, \quad \bm{P}^{-1} = \bm{J}^{t}\bm{P}^t\bm{J}, \quad \det\,\bm{P}(\sigma) = \det\,\bm{P}(0)  =  1.
\end{align}
Such a matrix can also be expressed in the direct form in terms of the ordered exponential (also known as the $\mathcal{T}$-exponential or path exponential):
\begin{align}
	& \bm{P}(\mathfrak{r}) = \mathcal{T}\exp\left\{\int_{0}^{\sigma}\bm{J} \bm{\nabla}_{\bm{f}}^2 \bm{H}(\sigma')d\sigma'\right\}.
\end{align}
From the definition of the propagator and Jacobi matrices we can see that 
\begin{align}
	& \bm{f}_{\mathfrak{r}}(\mathfrak{r}) =  \bm{P}(\mathfrak{r})\bm{f}_{\mathfrak{r}}(0,\mu).
\end{align}
Due to the fact, that Jacobi matrices have full rank (linearly independent columns), we can define Moore-Penrose (left) inverse matrices by the formula
\begin{align}
	& \bm{f}_{\mathfrak{r}}^{+} := \left(\bm{f}_{\mathfrak{r}}^t\bm{f}_{\mathfrak{r}}\right)^{-1}\bm{f}_{\mathfrak{r}}^t, &&  \bm{f}_{\mathfrak{r}}^{+} \bm{f}_{\mathfrak{r}} = \mathrm{I}_{3\times 3}.
\end{align}
Additionally we have the following relations between phase and ray coordinates:
\begin{align}
	& \delta \bm{f} = \bm{f}_{\mathfrak{r}} \delta \mathfrak{r}, && \delta \mathfrak{r} = \bm{f}_{\mathfrak{r}}^{+}\delta \bm{f},  \\
	&\bm{\nabla}_{\mathfrak{r}} \cdot = \bm{f}_{\mathfrak{r}}^t\bm{\nabla}_{\bm{f}}\cdot,&&
	\bm{\nabla}_{\bm{f}}\cdot =(\bm{f}_{\mathfrak{r}}^{+})^t\bm{\nabla}_{\mathfrak{r}}\cdot.
\end{align}
We can easily see that the derivatives $\bm{\nabla}_{\mathfrak{r}} \cdot$ depend on the choice of ray coordinates, but $\bm{\nabla}_{\bm{f}} \cdot$ is invariant under such changes. In the contrast to other gradients, $\bm{\nabla}_{\mathfrak{r}} \cdot$ can usually be computed along the reference ray. With this in mind, we will call  $\bm{\nabla}_{\bm{f}}\cdot$  \textit{observable} and $\bm{\nabla}_{\mathfrak{r}} \cdot$ \textit{computative} gradients, respectively. 

Let us discuss one more set variation equation, but now related to propagator matrix $\bm{P}$. Its variation satisfies the following system:
\begin{align}
	\label{second var syst}& \left\{\begin{array}{l}
		\frac{d }{d\sigma}\Delta[\bm{P}] =\bm{J} \bm{\nabla}_{\bm{f}}^2 \bm{H}\Delta[\bm{P}]+ \bm{J} \bm{\nabla}_{\bm{f}}^3 \bm{H}\{\bm{P},\bm{P}\}\Delta_0[\bm{f}] \\[1ex]
		\Delta[\bm{P}]|_{\sigma =0 }= \mathrm{0}_{6\times 6}
	\end{array}\right. ,
\end{align}
where
\begin{align}
	&   (\bm{\nabla}_{\bm{f}}^3\bm{H}\{\bm{P},\bm{P}\})_{ijk} = \left(\bm{\nabla}_{\bm{f}}^3\bm{H} \right)_{ilm} \bm{P}_{lj}\bm{P}_{mk}.
\end{align}
Then, introducing rank-3 propagation tensor $\mathfrak{P}$ as a solution of the system
\begin{align}
	& \left\{\begin{array}{l}
		\frac{d }{d\sigma}\mathfrak{P} =\bm{J} \bm{\nabla}_{\bm{f}}^2 \bm{H}\mathfrak{P}+ \bm{J} \bm{\nabla}_{\bm{f}}^3 \bm{H}\{\bm{P},\bm{P}\}  \\[1ex]
		\mathfrak{P}|_{\sigma =0 }= \mathrm{0}_{6\times 6\times 6}
	\end{array}\right. 
\end{align}
which again has explicit form
\begin{align}
	& \mathfrak{P}(\sigma) :=   \bm{P}(\sigma) \int_{0}^{\sigma} \bm{P}^{-1}(\sigma') \bm{J}\bm{\nabla}_{\bm{f}}^3\bm{H}\{\bm{P},\bm{P}\} (\sigma')  d\sigma',
\end{align}
we have expression for ray gradient of the propagator $\bm{P}$:
\begin{align}
	&  \bm{P}_{\mathfrak{r}}:=\bm{\nabla}_{\mathfrak{r}}\bm{P} = \mathfrak{P} \left\{\bm{f}_{\mathfrak{r}}(0,\mu),\cdot\right\}.
\end{align}

As it was mentioned earlier, variable $\sigma$ cannot have an explicit physical meaning. This raises the question: does anything in the considerations above change with the transition from $\sigma$ to some $\zeta$ (also a variable along the ray)? The answer is affirmative. Let the variable $\zeta$ be defined via the relation
\begin{align}
	& d \sigma :=\bm{s}(\bm{f}(\zeta)) d\zeta.
\end{align}
Then we have the Hamilton equation in the form
\begin{align}
	&         \frac{d}{d\zeta} \bm{f}(\zeta) =\bm{s}(\zeta) \bm{J} \bm{\nabla}_{\bm{f}}\bm{H}(\zeta),
\end{align}
and we see, that it differs only by the multiplicative factor $\bm{s}(\cdot)$.  However, for other equation we observe a significant change:
\begin{align}
	\frac{d }{d\zeta}\bm{P} &=\bm{J}[\bm{\nabla}_{\bm{f}}\bm{s}\bm{\nabla}_{\bm{f}}] \bm{H}\bm{P},\\
	\frac{d }{d\zeta}\mathfrak{P} &=\bm{J} [\bm{\nabla}_{\bm{f}}\bm{s}\bm{\nabla}_{\bm{f}}] \bm{H}\mathfrak{P} + \bm{J} [\bm{\nabla}_{\bm{f}}^2\bm{s}\bm{\nabla}_{\bm{f}}] \bm{H}\{\bm{P},\bm{P}\}.
\end{align}
This change is useful when we wish to transform from initial variable $\sigma$, defined by Hamiltonian itself, to more natural and appropriate one. Also, at this moment it appears obvious, that change $\mathcal{H}_{\mathrm{eff}}\to \bm{H}$, which was done before, did not impact on anything besides scaling of the natural variable $\sigma$.
\subsection{Transport equation}
We now consider the equation \ref{transport equation}. Let us rewrite it in more convenient form (omitting index $0$ for simplicity):
\begin{align}
	& \bm{\nabla}_{\bm{r}}\cdot\left\{A^2(\bm{r})\bm{k}(\bm{\nabla}\varphi(\bm{r}),\bm{r})\right\}=A^2(\bm{r})  g(\bm{\nabla}\varphi(\bm{r}),\bm{r}) \\ & g(\bm{\nabla}\varphi(\bm{r}),\bm{r}):= 2\left(\vec{\nabla}\varphi,\frac{\langle\vec{\nabla}\psi,\xi\rangle}{\langle\psi,\xi\rangle}\right) .
\end{align}
Using standard methods \cite{babich1998space,vcerveny2001seismic,kravtsov1990geometrical}, the solution can be expressed along the ray:
\begin{align}
	& A(\bm{r}(\sigma,\mu)) = A(\bm{r}(0,\mu))\times\sqrt{\frac{\det|\bm{r}_{\mathfrak{r}}(0,\mu)|}{\det|\bm{r}_{\mathfrak{r}}(\sigma,\mu)|}\exp\left\{\int_{0}^{\sigma} g(\sigma',\mu)d\sigma'\right\}},
\end{align}
where $\bm{r}_{\mathfrak{r}}(\sigma,\mu)$ is $3\times 3$ Jacobi matrix:
\begin{align}
	\bm{r}_{\mathfrak{r}}(\sigma,\mu)& = \bm{I}_{\bm{r}}\bm{f}_{\mathfrak{r}}(\sigma,\mu) = \bm{I}_{\bm{r}}\bm{P}(\sigma,\mu)\bm{f}_{\mathfrak{r}}(0,\mu),\quad
	\bm{I}_{\bm{r}}:= \left(\begin{array}{cc}
		\mathrm{I}_{3\times 3}&0\end{array}\right).
\end{align}
Then, we can write final form of the solution of the transport equation:
\begin{align}
	A(\bm{r}(\sigma,\mu))& = A(\bm{r}(0,\mu)) \left[\frac{\det \,|\bm{I}_{\bm{r}}\bm{f}_{\mathfrak{r}}(0,\mu)|}{\det \,|\bm{I}_{\bm{r}}\bm{P}(\sigma,\mu)\bm{f}_{\mathfrak{r}}(0,\mu)|}\right]^{\frac{1}{2}}\exp\mathfrak{T}(\sigma,\mu),\\
	\mathfrak{T}(\sigma,\mu)&:=\int\limits_{0}^\sigma \left(\vec{p}(\sigma',\mu),\frac{\langle\vec{\nabla}\psi,\xi\rangle}{\langle\psi,\xi\rangle}\right)d\sigma'.
\end{align}
\section{Dissipative case}
Let us recall that before we made the assumption \ref{assumption IM 0} that $\Im \lambda = 0$. Without this assumption, the problem may lack a well-defined solution. However, similar results can be obtained, if we assume that imaginary parts of eigenvalues are \textit{small enough}:
\begin{align}
	& \lambda (p_{\tau},\vec{r}) = \bm{\lambda} (p_{\tau},\vec{r}) + i \varepsilon \tilde{\lambda}(p_{\tau},\vec{r}),
\end{align}
where both $\bm{\lambda}$ and $\tilde{\lambda}$ are real-valued. Then, we would need to change Hamiltonian $\mathcal{H}_{\mathrm{eff}}$:
\begin{align}
	\mathcal{H}_{\mathrm{eff}}(\bm{p},\bm{r}) &= \left[p_\tau^2+ \bm{\lambda}(p_{\tau},\vec{r}) - p_x^2-p_y^2\right] + i\varepsilon \tilde{\lambda}(p_{\tau},\vec{r}),\\
	\bm{H}(\bm{p},\bm{r})	&: = \frac{p_\tau^2+\bm{\lambda}(p_{\tau},\vec{r}) - p_x^2-p_y^2}{2}.
\end{align} 
Then, corresponding eikonal and transport equations (reflecting orders $\varepsilon^0$ and $\varepsilon^1$ in the asymptotic decomposition) change form to:
\begin{align}
	\label{eikonal equation diss} \bm{H}(\bm{\nabla}\varphi(\bm{r}),\bm{r}) &= 0,\\
	\label{transport equation diss} \bm{\nabla}_{\bm{r}}\cdot\left\{[A^0(\bm{r})]^2\bm{k}(\bm{\nabla}\varphi(\bm{r}),\bm{r})\right\}&=2[ A^0 (\bm{r})]^2\left[\left(\vec{\nabla}\varphi,\frac{\langle\vec{\nabla} \psi^0,\xi^0\rangle}{\langle\psi^0,\xi^0\rangle}\right) + \tilde{\lambda}\right],
\end{align}
while $\bm{k}$ is defined by $\bm{H}$. Since these equations resemble those in the previous section, we summarize their solutions as follows:
\begin{itemize}
	\item Trajectories $\bm{f}(\sigma)$, as well as phase $\varphi$ itself are defined by  \textit{real-valued} Hamiltonian $\bm{H} = \frac{1}{2}\Re\mathcal{H}_{\mathrm{eff}} $: 
	\begin{align}
		\label{phase new} \varphi(\sigma,\mu) &= \varphi_0(\mu) + \int\limits_{0}^{\sigma}p_{\tau}(\sigma',\mu)\left[\frac{1}{2}\frac{\partial \bm{\lambda}}{\partial p_{\tau}}(\sigma',\mu) - \frac{\bm{\lambda}(\sigma',\mu)}{p_{\tau}(\sigma',\mu)}\right]d\sigma'  .
	\end{align}
	\item Amplitude now includes additional exponential factor, defined by the presence of dissipation in system:
	\begin{align}
		A(\bm{r}(\sigma,\mu))& = A(\bm{r}(0,\mu)) \left[\frac{\det \,|\bm{I}_{\bm{r}}\bm{f}_{\mathfrak{r}}(0,\mu)|}{\det \,|\bm{I}_{\bm{r}}\bm{P}(\sigma,\mu)\bm{f}_{\mathfrak{r}}(0,\mu)|}\right]^{\frac{1}{2}}\exp\mathfrak{T}_{\mathrm{diss}}(\sigma,\mu),\\
		\mathfrak{T}_{\mathrm{diss}}(\sigma,\mu)&:=\int\limits_{0}^\sigma \left[\left(\vec{p}(\sigma',\mu),\frac{\langle\vec{\nabla}\psi,\xi\rangle}{\langle\psi,\xi\rangle}\right)+ \tilde{\lambda}(\sigma',\mu)\right]d\sigma'.
	\end{align}
\end{itemize}

\section{Fronts and observations}
\subsection{General idea}
Consider an arbitrary function $g$ along a ray given by:
\begin{align}
	& g(\bm{f}) = g(\bm{f}(\mathfrak{r})).
\end{align}
For such function we would be interested in two objects - \textit{ray} ($\bm{\nabla}_{\mathfrak{r}}$) and \textit{observable} (space-time,  $\bm{\nabla}_{\bm{r}}$) gradients and relations between them. From the previous considerations we know, that
\begin{align}
	&\bm{\nabla}_{\bm{r}}g(\bm{f}) = \bm{I}_{\bm{r}}(\bm{f}_{\mathfrak{r}}^{+})^t\bm{\nabla}_{\mathfrak{r}}g(\bm{f}(\mathfrak{r}))
\end{align}
Then, surfaces of $g(\bm{f})= \rm{const}$ can be (locally) described using such gradients, obviously, outside the caustics, where we do not have bijective relation between $\bm{r}$ and $\mathfrak{r}$. Before analyzing specific functions and their fronts, we perform a natural change of variable  $\sigma$ along with introduction of additional notations:
\begin{align}
	& d\bm{\tau} := \frac{\partial \bm{H}}{\partial p_{\tau}} d\sigma = \left[p_{\tau} + \frac{1}{2}\frac{\partial \bm{\lambda}}{\partial p_{\tau}}\right]d\sigma,\quad \bm{\tau}(0) = 0. 
\end{align}
we have the following form of the Hamiltonian system:
\begin{align}
	&  \frac{d}{d\bm{\tau}}\left(\begin{array}{c}
		\tau\\
		\vec{r}\\
		p_{\tau}\\
		\vec{p}
	\end{array}\right) = \left(\begin{array}{c}
		1\\
		\frac{\vec{p}}{p_{\tau} + \frac{1}{2}\frac{\partial \bm{\lambda}}{\partial p_{\tau}}}\\
		0\\
		-\frac{1}{2}\frac{\vec{\nabla}k^2}{p_{\tau} + \frac{1}{2}\frac{\partial \bm{\lambda}}{\partial p_{\tau}}}
	\end{array}\right). 
\end{align}
From here we can see, that value 
\begin{align}
	& \vec{v}:=\frac{\vec{p}}{p_{\tau} + \frac{1}{2}\frac{\partial \bm{\lambda}}{\partial p_{\tau}}} = \frac{\sqrt{p_{\tau}^2 + \bm{\lambda}(p_{\tau},\vec{r})}}{p_{\tau} + \frac{1}{2}\frac{\partial \bm{\lambda}}{\partial p_{\tau}}}\frac{\vec{p}}{|\vec{p}|}=\left[\frac{\partial}{\partial p_{\tau}}\sqrt{p_{\tau}^2 + \bm{\lambda}(p_{\tau},\vec{r})}\right]^{-1} \frac{\vec{p}}{|\vec{p}|}
\end{align}
plays the role of group speed for the system, while equations for time parts ($\tau,p_{\tau}$) provide us simple solutions:
\begin{align}
	\tau(\bm{\tau},\mu)   = \tau_0(\mu) + \bm{\tau}, &&p_\tau(\bm{\tau},\mu)   = p_\tau(0,\mu).
\end{align} 
With the equations expressed in natural variables, we now compute gradients for different functions.

Additionally, before exact examples of functions $g$, let us study the following case:
\begin{align}
	g(\bm{f}(\mathfrak{r})) = \int_0^{\bm{\tau}} G(\bm{f}(\bm{\tau}',\mu))d\bm{\tau}'.
\end{align}
Then we have the following formula for ray gradient:
\begin{align}
	\bm{\nabla}_{\mathfrak{r}} g(\bm{f}(\bm{\tau},\mu)) &= \left(\begin{array}{c}
		G(\bm{f}(\bm{\tau},\mu))\\
		\nabla_{\mu}^{\mathrm{int}}[G](\bm{f}(\bm{\tau},\mu))
	\end{array}\right),\\
	\nabla_{\mu}^{\mathrm{int}}[G](\bm{f}(\bm{\tau},\mu)) & := \bm{I}_{\mu} \bm{f}_{\mathfrak{r}}^t(0,\mu)\int_0^{\bm{\tau}} \bm{P}^t(\bm{\tau}',\mu) \bm{\nabla}_{\bm{f}} G(\bm{f}(\bm{\tau}',\mu))d\bm{\tau}',\\
	\bm{I}_{\mu} & :=\left(\begin{array}{ccc}
		0 & 1 & 0\\
		0 & 0 & 1
	\end{array}\right).
\end{align}

\subsection{Time, distance, phase}
In this section, we would discuss 4 simplest examples of functions $g$:
\begin{align}
	\tau(\bm{f}(\mathfrak{r}))&:= \tau_0(\mu) + \bm{\tau},\quad \bm{\tau}(\bm{f}(\mathfrak{r})):= \bm{\tau}, \quad \bm{l}(\bm{f}(\mathfrak{r})):= \int_{0}^{\bm{\tau}}|\vec{v}|d\bm{\tau}',\\
	\varphi(\bm{f}(\mathfrak{r})) &=    \varphi_0(\mu) + p_{\tau}\bm{\tau}-\int_0^{\bm{\tau}}|\vec{v}||\vec{p}|d\bm{\tau}'.
\end{align}
Ray gradients of this functions have the following form:
\begin{align}
	\bm{\nabla}_{\mathfrak{r}} \tau&:= \left(\begin{array}{c}
		1\\
		\nabla_{\mu}\tau_0
	\end{array}\right),\quad
	\bm{\nabla}_{\mathfrak{r}} \bm{\tau}:= \left(\begin{array}{c}
		1\\
		0
	\end{array}\right),\quad
	\bm{\nabla}_{\mathfrak{r}} \bm{l}:= \left(\begin{array}{c}
		|\vec{v}|\\
		\nabla_{\mu}^{\mathrm{int}}[|\vec{v}|]
	\end{array}\right),\\
	\bm{\nabla}_{\mathfrak{r}} \varphi &:= \left(\begin{array}{c}
		p_{\tau} - |\vec{v}||\vec{p}|\\
		\nabla_{\mu}\varphi_0 +\bm{\tau}\nabla_{\mu}p_{\tau}(0,\mu) - \nabla_{\mu}^{\mathrm{int}}[|\vec{v}| |\vec{p}|]
	\end{array}\right).
\end{align}
\subsection{Amplitude}
We now treat the amplitude $A$ as the function $g$. From the calculations above we can see, that it has more complex expression on ray, which cannot be written as simple integral of some function along it. For gradient of amplitude we would use the following formula:
\begin{align}
	\notag&\bm{\nabla}_{\mathfrak{r}} A(\bm{\tau},\mu) =  A(\bm{\tau},\mu)\times\bigg\{ A^{-1}(0,\mu) \bm{\nabla}_{\mathfrak{r}} A(0,\mu) + \bm{\nabla}_{\mathfrak{r}} \mathfrak{T}_{\mathrm{diss}}(\bm{\tau},\mu)- \\&- \frac{1}{2} \bm{\nabla}_{\mathfrak{r}}\ln\det \,|\bm{I}_{\bm{r}}\bm{P}(\bm{\tau}',\mu)\bm{f}_{\mathfrak{r}}(0,\mu)|\bigg|_{0}^{\bm{\tau}}\bigg\}.
\end{align}   
Each part should be treated separately. Let us start from initial amplitude $A(\bm{r}(0,\mu))$. Unlike initial phase $\varphi_0$, which has to satisfy some conditions as a solution of Hamilton system, we can define it arbitrary. Natural way is to define it as function of $\bm{r}=\bm{r}_0(\mu)$ and write derivative in the form:
\begin{align}
	\bm{\nabla}_{\mathfrak{r}} A(\bm{r}(0,\mu)) & = \bm{r}_{\mathfrak{r}}^t(0,\mu) \bm{\nabla_{\bm{r}}}A(\bm{r}(0,\mu)).
\end{align}
For integral part we have:
\begin{align}
	\bm{\nabla}_{\mathfrak{r}} \mathfrak{T}_{\mathrm{diss}}(\bm{\tau},\mu):= \left(\begin{array}{c}
		\left(\vec{v} ,\frac{\langle\vec{\nabla} \psi,\xi\rangle}{\langle\psi,\xi\rangle}\right)+\frac{\tilde{\lambda}}{p_{\tau} + \frac{1}{2}\frac{\partial \bm{\lambda}}{\partial p_{\tau}}}\\
		\nabla_{\mu}^{\mathrm{int}}\left[\left(\vec{v} ,\frac{\langle\vec{\nabla} \psi,\xi\rangle}{\langle\psi,\xi\rangle}\right)+\frac{\tilde{\lambda}}{p_{\tau} + \frac{1}{2}\frac{\partial \bm{\lambda}}{\partial p_{\tau}}}\right]
	\end{array}\right).
\end{align}
Now, let us consider determinant part. Key to the computation would be the following formula:
\begin{align}
	& \frac{d}{ds}\ln \det |M(s)| =  \mathrm{tr}\left(M^{-1}\frac{dM}{ds}\right),
\end{align}
which is valid for all invertable $M$. Such result allows us to have
\begin{align}
	\notag\bm{\nabla}_{\mathfrak{r}}\ln\det \,|\bm{I}_{\bm{r}}\bm{P}(\bm{\tau},\mu)\bm{f}_{\mathfrak{r}}(0,\mu)| =\mathrm{tr} \left([\bm{I}_{\bm{r}}\bm{P}\bm{f}_{\mathfrak{r}}(0)]^{-1}\bm{I}_{\bm{r}}\left[\bm{P}\bm{\nabla}_{\mathfrak{r}}\bm{f}_{\mathfrak{r}}(0)+\bm{\nabla}_{\mathfrak{r}}\bm{P}\bm{f}_{\mathfrak{r}}(0)\right]\right),
\end{align}
or, in more explicit form,
\begin{align}
	\notag\bm{\nabla}_{\mathfrak{r}}&\ln\det \,|\bm{I}_{\bm{r}}\bm{P}(\bm{\tau},\mu)\bm{f}_{\mathfrak{r}}(0,\mu)| =\\&= \mathrm{tr} \left([\bm{I}_{\bm{r}}\bm{P}(\bm{\tau})\bm{f}_{\mathfrak{r}}(0)]^{-1}\bm{I}_{\bm{r}}\left[\bm{P}(\bm{\tau})\bm{f}_{\mathfrak{r}\mathfrak{r}}(0)+\mathfrak{P}(\bm{\tau})\left\{\bm{f}_{\mathfrak{r}}(0), \bm{f}_{\mathfrak{r}}(0)\right\}\right]\right).
\end{align}

\section{Example}
\subsection{Simple model}
Consider a simplified case that permits exact solutions for the vertical modes:
\begin{align}
	\nu^2(\vec{r},z)&\equiv \bm{\nu}^2   = \mathrm{const},\quad 0\le z\le\bm{h}(\vec{r}),\\
	\bm{w}^2 (p_{\tau},\vec{r})&= \bm{\nu}^2 p_{\tau}^2 \bm{h}^2(\vec{r}).
\end{align}
Then we have simple formulas for  eigenfunctions
\begin{align}
	\bm{\psi} (s,k) & =\frac{\sin( \sqrt{\bm{w}^2 - k^2} s)}{\sin( \sqrt{\bm{w}^2 - k^2} )},\quad
	||\bm{\psi}||^2  = \frac{1}{2}  \left[ 1 -\frac{\sin( 2\sqrt{\bm{w}^2 - k^2} )}{2 \sqrt{\bm{w}^2 - k^2} } \right],
\end{align}
while eigenvalues $k$ are solutions of the equation
\begin{align}
	\cot ( \sqrt{\bm{w}^2 - k^2} ) = -\frac{\alpha k}{\sqrt{\bm{w}^2 - k^2}}.
\end{align}
Such solutions can be approximated in terms of dependence on parameter $\alpha$:
\begin{align}
	k^2_l(\alpha) & \approx k^2_l(0)\left[1 - \frac{2 \alpha}{k_l(0) + \alpha \frac{\bm{w}^2}{q_l^2}} \right]= \bm{w}^2 - q_l^2 \left[1 + \frac{2 \alpha}{k_l(0) + \alpha \frac{\bm{w}^2}{q_l^2}}\frac{k_l^2(0)}{q_l^2}\right],\\ k_l^2(0)&:= \bm{w}^2 - q_l^2,\quad q_l:= \pi \left(l + \frac{1}{2}\right).
\end{align}
Eigenvalues of the general problem are related to ones above via
\begin{align}
	\lambda_l(p_{\tau},\vec{r},\alpha) := \frac{k_l^2(p_{\tau},\vec{r},\alpha) }{\bm{h}^2(\vec{r})},\quad \lambda_l(p_{\tau},\vec{r},0) =  \bm{\nu}^2 p_{\tau}^2  - \frac{q_l^2}{\bm{h}^2(\vec{r})}.
\end{align}
Note that these relations describe only the real eigenvalues and corresponding eigenfunctions in the area, where $k_l^2(0)>0$, that are our main focus. For complex eigenvalues, corresponding to leaking modes equations above are not exactly correct. 

Other results for this case follow from these exact (for $\alpha = 0$) or approximate expressions. Additionally we can see, that since dependence of all eigenvalues and eigenfunctions on the horizontal coordinates $\vec{r}$ is through function $\bm{h}(\vec{r})$, we have the following form for the integral function in the exponential factor for amplitude:
\begin{align}
	\mathfrak{T}(\bm{\tau},\mu)  = (1-\alpha)\int_{0}^{\bm{\tau}}\theta_l(\vec{r}(\bm{\tau}',\mu),p_{\tau}(\mu),\alpha) \left(\vec{p}(\bm{\tau},\mu),\frac{\vec{\nabla}\bm{h}(\vec{a}(\bm{\tau}',\mu))}{\bm{h}(\vec{a}(\bm{\tau}',\mu))}\right)d\bm{\tau}',
\end{align}
where $\theta_l$ is some function, which exact expression is too complex and unnecessary to be provided here.
\subsection{Numerical example}
As an illustration to the simple case above, we would present some numerical examples. We would compare 3 situations:
\begin{itemize}
	\item "ideal" (Neuman) case: $\alpha = 0$,
	\item self-adjoint case: $\alpha = 1$,
	\item example of general case: $\alpha = \frac{1}{2}$.
\end{itemize}
For all cases, the medium parameters are:
\begin{align}
	& \bm{c}= 1500,\quad \mathrm{c}_{\mathrm{bot}} = 1700,\quad \bm{\nu}^2 = \left(\frac{8}{15}\right)^2,\quad \bm{h}(\vec{r}) = 10 + 10^{-3} x,
\end{align}
while source would be described using formulas:
\begin{align}
	&	\tau_0(\mu_1,\mu_2) = \mathrm{c}_{\mathrm{bot}}\mu_1, && p_{\tau}(\mu_1,\mu_2) = - \frac{2\pi  }{\mathrm{c}_{\mathrm{bot}}}(300 + 50 \mu_1)  ,\\
	&	\vec{r}(\mu_1,\mu_2) = \left(\begin{array}{c}
		\cos\mu_2\\
		\sin \mu_2
	\end{array}\right),&& \vec{p}(\mu_1,\mu_2) =\frac{k_l(\mu_1,\mu_2)}{\bm{h}(\mu_1,\mu_2)}\left(\begin{array}{c}
		\cos\mu_2\\
		\sin \mu_2
	\end{array}\right).
\end{align}
For such source we would show most interesting comparisons of time,amplitude and phase fronts for different combinations of parameters.

\begin{figure}
	\centering
	\begin{subfigure}{.33\textwidth}
		\centering
		\includegraphics[height=1.6\linewidth]{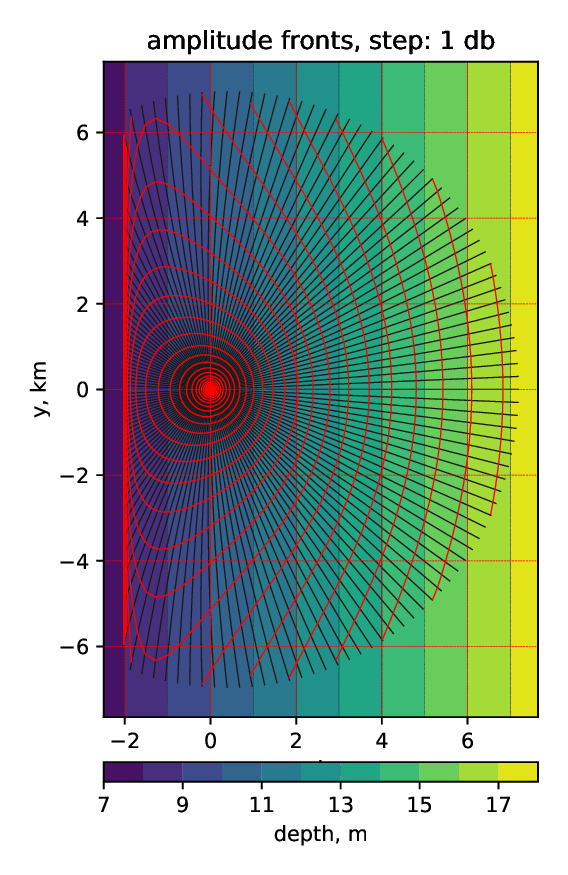}
		\caption{ $\alpha = 0$}
		\label{afa0}
	\end{subfigure}%
	\begin{subfigure}{.33\textwidth}
		\centering
		\includegraphics[height=1.6\linewidth]{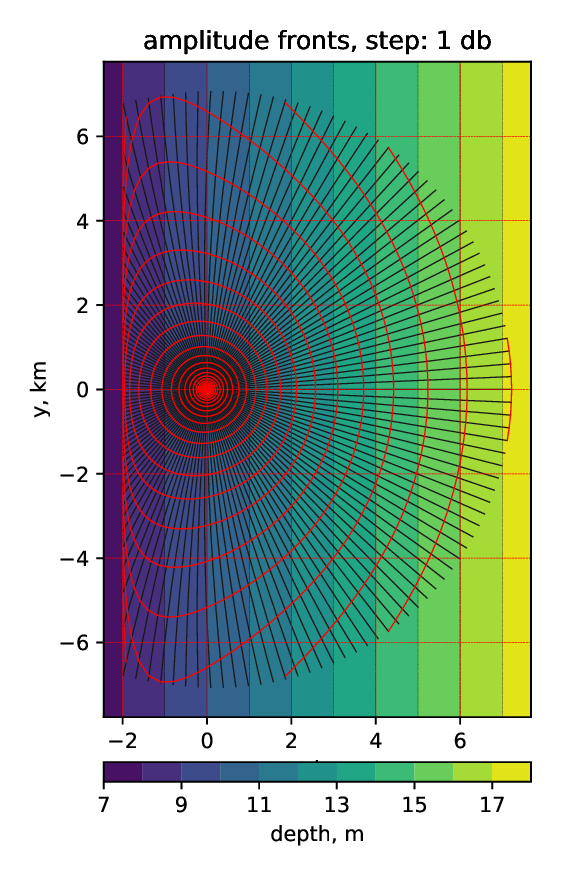}
		\caption{$\alpha = \frac{1}{2}$}
		\label{afa12}
	\end{subfigure}
	\begin{subfigure}{.33\textwidth}
		\centering
		\includegraphics[height=1.6\linewidth]{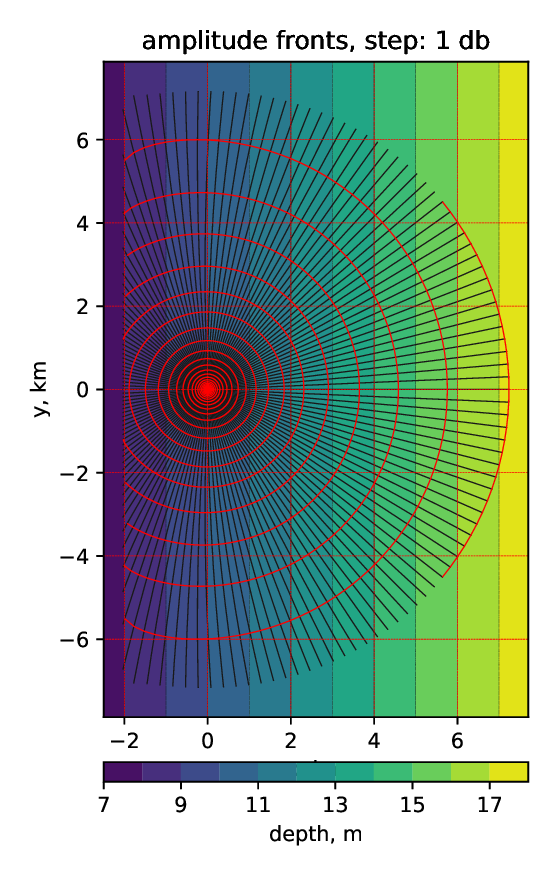}
		\caption{$\alpha = 1$}
		\label{afa1}
	\end{subfigure}
	\caption{Amplitude fronts, $l = 1$, $\mu_1 = 0$, $\mu_2\in[-\pi,\pi]$, $\bm{\tau} = 5$}
	\label{af}
\end{figure}
On the Figure \ref{af} we see, that for the same pulse, change in the parameter $\alpha$ significantly changes amplitude distribution along rays. For $\alpha = 1$ (\ref{afa1}) we have approximately zero difference to amplitude fronts we expect in the free space - just spherical decay, while for $\alpha = \frac{1}{2}$ (\ref{afa12}) and even $\alpha = 0$ (\ref{afa0}) we have existence of the exponential decay near the critical depth, such that $k_l^2(0) = 0$, defined \textit{purely} by the structure of eigenfunctions.
\begin{figure}
	\centering
	\begin{subfigure}{.33\textwidth}
		\centering
		\includegraphics[height=1.6\linewidth]{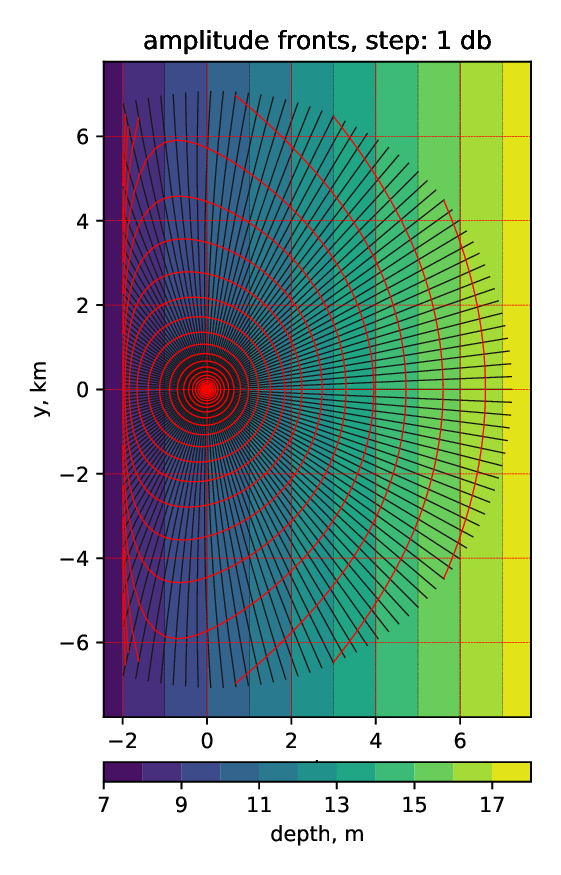}
		\caption{ $\mu_1 = 0$}
		\label{afmu0}
	\end{subfigure}%
	\begin{subfigure}{.33\textwidth}
		\centering
		\includegraphics[height=1.6\linewidth]{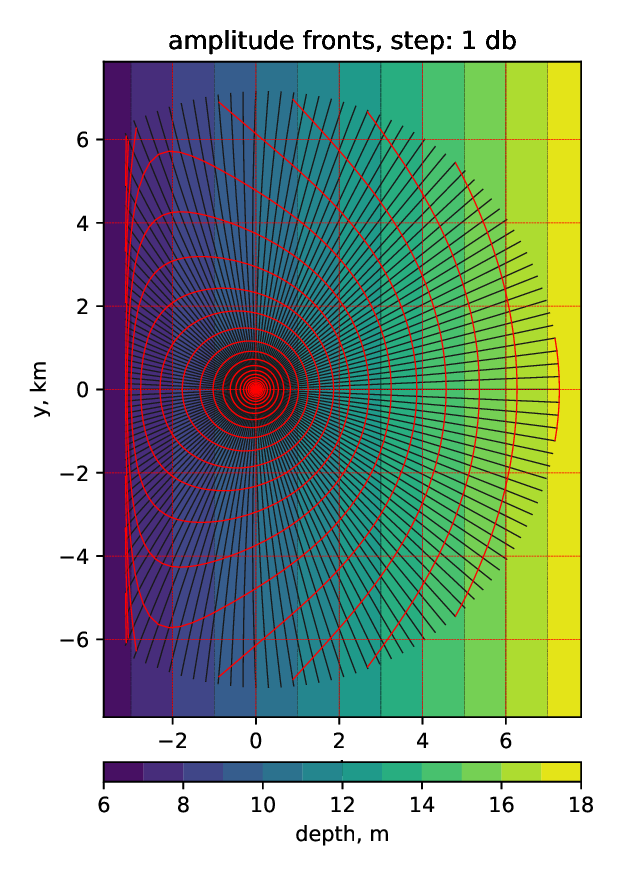}
		\caption{$\mu_1 =1$}
		\label{afmu1}
	\end{subfigure}
	\begin{subfigure}{.33\textwidth}
		\centering
		\includegraphics[height=1.6\linewidth]{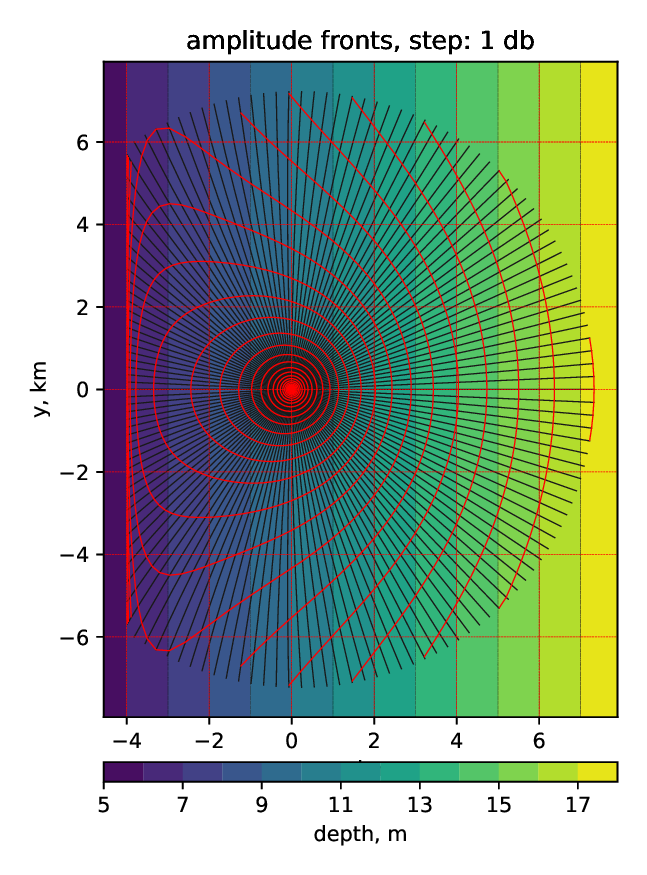}
		\caption{$\mu_1 = 2$}
		\label{afmu2}
	\end{subfigure}
	\caption{Amplitude fronts, $\alpha = \frac{1}{2}$, $l = 1$, $\mu_2\in[-\pi,\pi]$, $\bm{\tau} = 5$}
	\label{afmu}
\end{figure}

Next, on the Figure \ref{afmu} we see, that change of the frequency (defined by the parameter $\mu_1$: from 300 Hz at $\mu_1=0$ (\ref{afmu0}) to 400 Hz at $\mu_1 = 2$ (\ref{afmu2})) also significantly impacts on the whole amplitude distribution.
\begin{figure}
	\centering
	\begin{subfigure}{.33\textwidth}
		\centering
		\includegraphics[height=2\linewidth]{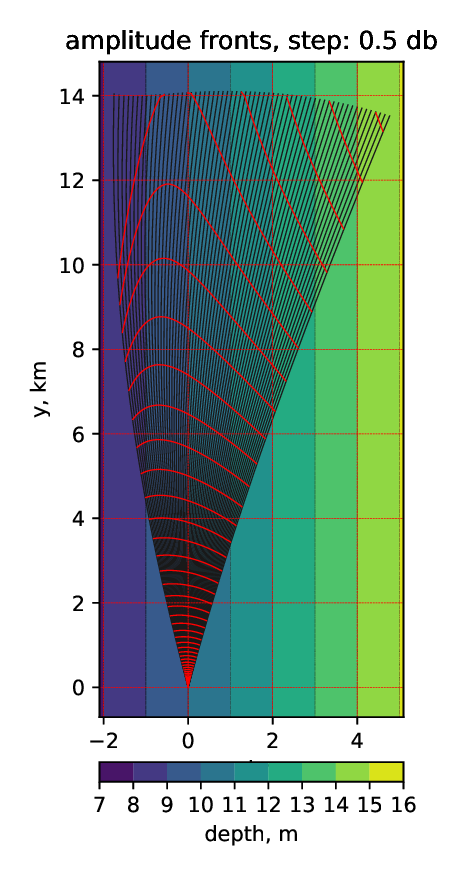}
		\caption{ $\mu_1 = 0$}
		\label{afmu0sec}
	\end{subfigure}%
	\begin{subfigure}{.33\textwidth}
		\centering
		\includegraphics[height=2\linewidth]{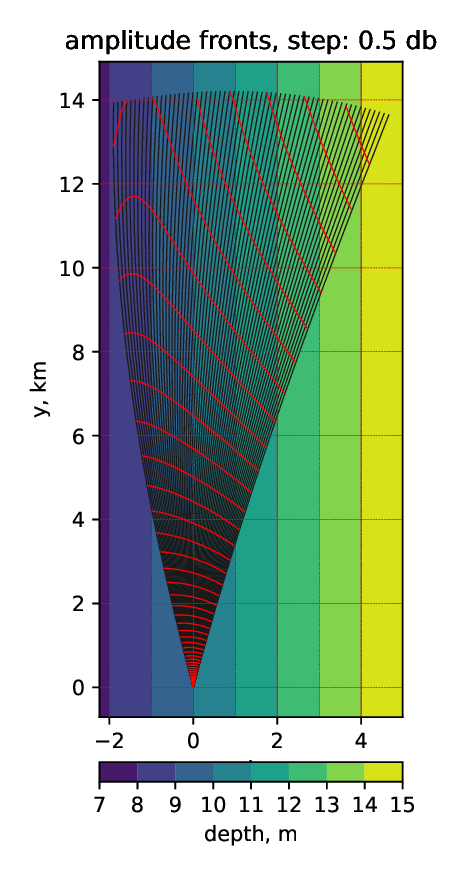}
		\caption{$\mu_1 =0.5$}
		\label{afmu1sec}
	\end{subfigure}
	\begin{subfigure}{.33\textwidth}
		\centering
		\includegraphics[height=2\linewidth]{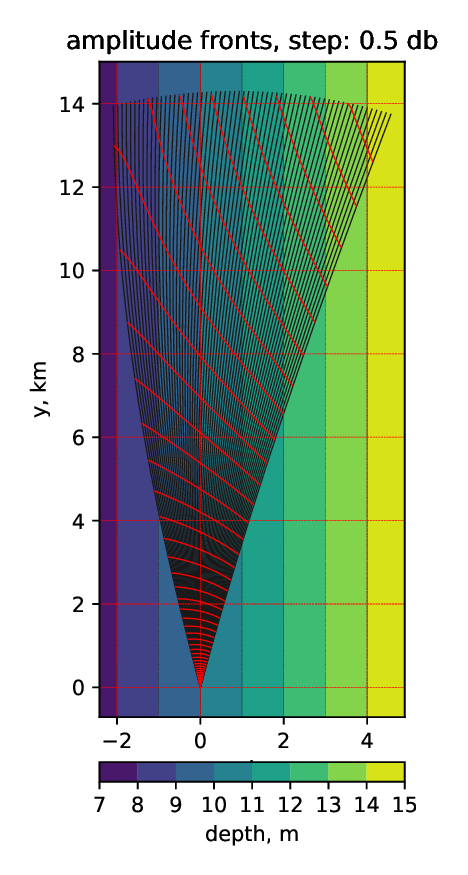}
		\caption{$\mu_1 = 1$}
		\label{afmu2sec}
	\end{subfigure}
	\caption{Amplitude fronts, $\alpha = \frac{1}{2}$, $l = 1$, $\mu_2\in\left[\frac{5}{12}\pi  ,\frac{7}{12}\pi \right]$, $\bm{\tau} = 10	$}
	\label{afmusec}
\end{figure}
\begin{figure}
	\centering
	\begin{subfigure}{.33\textwidth}
		\centering
		\includegraphics[height=2\linewidth]{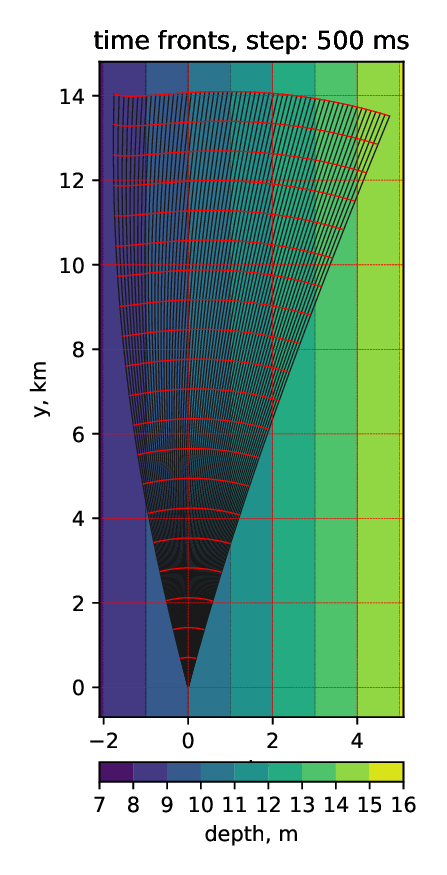}
		\caption{ $\mu_1 = 0$}
		\label{tfmu0sec}
	\end{subfigure}%
	\begin{subfigure}{.33\textwidth}
		\centering
		\includegraphics[height=2\linewidth]{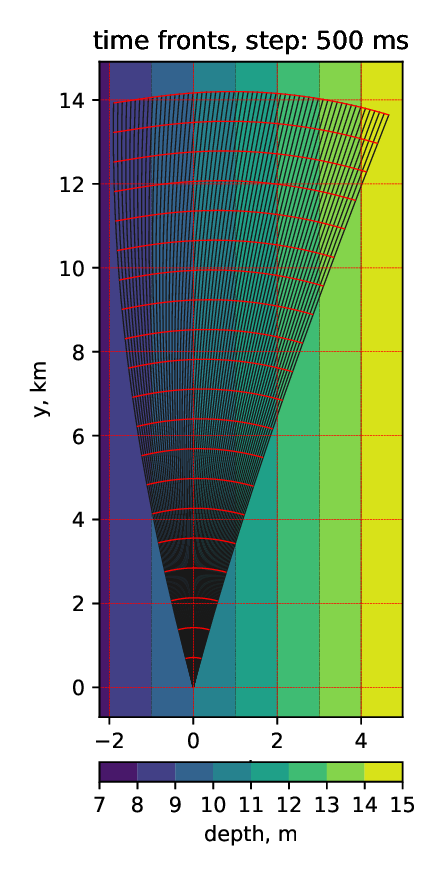}
		\caption{$\mu_1 =0.5$}
		\label{tfmu1sec}
	\end{subfigure}
	\begin{subfigure}{.33\textwidth}
		\centering
		\includegraphics[height=2\linewidth]{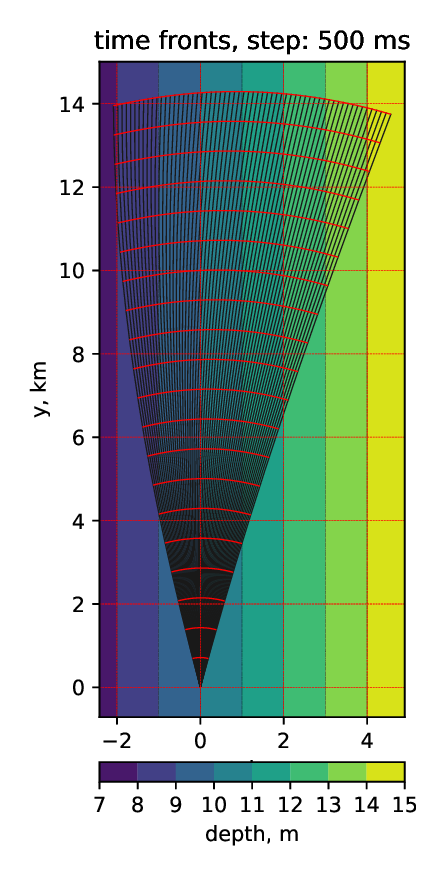}
		\caption{$\mu_1 = 1$}
		\label{tfmu2sec}
	\end{subfigure}
	\caption{Time fronts, $\alpha = \frac{1}{2}$, $l = 1$, $\mu_2\in\left[\frac{5}{12}\pi  ,\frac{7}{12}\pi \right]$, $\bm{\tau} = 10	$}
	\label{tfmusec}
\end{figure}

On the last two Figures \ref{afmusec} and \ref{tfmusec} we can see the more detailed version of the previous result, where we can see the change not only in amplitude fronts itself (\ref{afmu0sec} to \ref{afmu2sec}), but also in the form of the pulse itself - rays for different frequencies but same time steps have obvious difference one from another.

\section{Conclusion}

The present study extends the theoretical foundations of acoustic pulse propagation in shallow-water environments by incorporating non-self-adjoint operators and $\varepsilon$-pseudodifferential formulations. This approach captures essential non-ideal effects such as bottom interaction, mode leakage, and geometric dispersion?phenomena critical for realistic modeling of underwater acoustic fields.  

Future research should focus on several key directions.  
First, the present framework can be expanded to \emph{multi-mode coupling}, allowing for a more complete description of broadband acoustic fields and mode interference patterns. Such generalizations would bridge the gap between analytical models and full numerical simulations, as suggested in recent works on adiabatic and nonadiabatic mode interactions~\cite{jensen2011computational, porter1987gaussian}.

Second, \emph{numerical implementation and comparison with field data} will be crucial to validate the theoretical predictions derived from the $\varepsilon$-PDO formalism. The development of hybrid models combining ray-based Hamiltonian dynamics with parabolic equation solvers or normal-mode codes (e.g., KRAKEN~\cite{collins1993splitstep}) could significantly improve computational efficiency and accuracy in coastal acoustics applications.  

Finally, the general mathematical methods developed here --- such as Hamiltonian variation analysis \cite{farra1993ray,farra1989ray} and asymptotic mode reconstruction --- have potential beyond underwater acoustics. They may be applicable to a wide class of \emph{wave propagation problems in layered or dissipative media}, including seismology, optics, and plasma physics, where non-self-adjoint operators and geometric dispersion play similar roles~\cite{vcerveny2001seismic, chapman2004fundamentals}.  

By linking semiclassical and asymptotic approaches with modern computational methods, this research provides a path toward a unified mathematical framework for analyzing acoustic pulse propagation in complex, non-ideal oceanic environments.

\bibliographystyle{plain}
\bibliography{sn-bibliography}

\end{document}